\documentclass[a4paper,twocolumn]{article}

\usepackage{moreverb,url}
\usepackage{amsmath,amsfonts}
\usepackage[round]{natbib}
\usepackage{booktabs}
\usepackage{titlesec}
\usepackage[font=small,labelfont={sf,bf},textfont={sf}]{caption}

\usepackage[colorlinks,bookmarksopen,bookmarksnumbered,allcolors=black]{hyperref}

\titleformat{\section}
  {\normalfont\sffamily\bfseries\large}
  {\thesection}{10pt}{}
\titleformat{\subsection}
  {\normalfont\sffamily\bfseries\large}
  {\thesubsection}{5pt}{}

\newcommand\BibTeX{{\rmfamily B\kern-.05em \textsc{i\kern-.025em b}\kern-.08em
T\kern-.1667em\lower.7ex\hbox{E}\kern-.125emX}}

\usepackage{pgfplots}
\usepackage{pgfplotstable}
\usepackage[super]{nth}
\usepackage{algorithmicx}
\usepackage{algpseudocode}
\usepackage{algorithm}
\usepackage[eulergreek]{sansmath}

\pgfplotsset{compat=1.11,
        /pgfplots/ybar legend/.style={
        /pgfplots/legend image code/.code={%
        \draw[##1,/tikz/.cd,bar width=3pt,yshift=-0.2em,bar shift=0pt]
                plot coordinates {(0cm,0.8em)};},
},
tick label style = {font=\sansmath\sffamily\footnotesize},
legend style={font=\footnotesize\sansmath\sffamily},
every axis label/.append style={font=\sffamily\footnotesize},
}

\usetikzlibrary{decorations.pathreplacing,calc, matrix, backgrounds,
arrows.meta, matrix, fit}
\tikzset{%
  highlight/.style={rectangle,rounded corners,fill=red!15,draw,fill opacity=0.5,thick,inner sep=0pt}
}

\newcommand{\tikzmark}[1]{\tikz[overlay,remember picture] \node (#1) {};}
\newcommand{\myvector}[1]{\boldsymbol{ \mathbf #1 }}

\usepackage{soul}

\tikzset{
  dim above/.style={to path={\pgfextra{
        \pgfinterruptpath
        \draw[>=latex,|<->|] let
        \p1=($(\tikztostart)!10mm!90:(\tikztotarget)$),
        \p2=($(\tikztotarget)!10mm!-90:(\tikztostart)$)
        in(\p1) -- (\p2) node[pos=.5,sloped,above]{#1};
        \endpgfinterruptpath
      }(\tikztostart) -- (\tikztotarget) \tikztonodes
    }
  },
  dim below/.style={to path={\pgfextra{
        \pgfinterruptpath
        \draw[>=latex,|<->|] let
        \p1=($(\tikztostart)!10mm!90:(\tikztotarget)$),
        \p2=($(\tikztotarget)!10mm!-90:(\tikztostart)$)
        in (\p1) -- (\p2) node[pos=.5,sloped,aobve]{#1};
        \endpgfinterruptpath
      }(\tikztostart) -- (\tikztotarget) \tikztonodes
    }
  },
}

\newcommand{\norm}[1]{\left\lVert#1\right\rVert}

\newcommand*{\AddLightNote}[4]{%
    \begin{tikzpicture}[overlay, remember picture]
        \draw [decoration={amplitude=0.5em},decorate,thick,gray]
            ($(#3)!(#1.north)!($(#3)-(0,1)$)$) --
            ($(#3)!(#2.south)!($(#3)-(0,1)$)$)
                node [align=center, text width=2.5cm, pos=0.5, anchor=west] {#4};
    \end{tikzpicture}
}%

\newcommand*{\AddOneLineNote}[4]{%
    \begin{tikzpicture}[overlay, remember picture]
        \draw [decoration={amplitude=0.5em},decorate,thick,gray]
          ($(#3)!(#1.north)!($(#3)-(0,1)$)$) --
          ($(#3)!(#2.south)!($(#3)-(0,1)$)$)
          node [align=center, text width=2.5cm, pos=0.5, anchor=west] {#4};
    \end{tikzpicture}
}

\algnewcommand{\algorithmicgoto}{\textbf{go to}}%
\algnewcommand{\Goto}[1]{\algorithmicgoto~\ref{#1}}%

\usetikzlibrary{positioning}
\usetikzlibrary{patterns}
\makeatletter
\definecolor{gnuplot@orange}{RGB}{229,158,0}
\definecolor{gnuplot@purple}{RGB}{148,0,212}
\definecolor{gnuplot@red}{RGB}{200,0,0}
\definecolor{gnuplot@lightblue}{RGB}{87,181,232}
\definecolor{gnuplot@green}{RGB}{0,158,65}
\definecolor{gnuplot@darkblue}{RGB}{0,95,169}
\definecolor{gnuplot@yellow}{RGB}{240,227,66}
\pgfplotscreateplotcyclelist{colorGPL}{%
gnuplot@darkblue,every mark/.append style={fill=gnuplot@darkblue!80!black},mark=o\\%
gnuplot@red,every mark/.append style={fill=gnuplot@red!50!black},mark=square\\%
gnuplot@green,every mark/.append style={fill=gnuplot@green!50!black},mark=triangle*\\%
gnuplot@orange,every mark/.append style={fill=gnuplot@orange!80!black,mark size=2.5pt},mark=x\\%
gnuplot@purple,mark=diamond\\%
black,every mark/.append style={solid,fill=gnuplot@darkblue!80!black},mark=square*\\%
gnuplot@lightblue,every mark/.append style={solid,fill=gnuplot@lightblue!30!black},mark=otimes*\\%
red!80!white,densely dashed,every mark/.append style={solid,fill=red!40!white},mark=oplus*\\%
}
\makeatother

\pgfplotsset{compat=1.9}

\newif\ifnotanon
\notanontrue

\begin{document}

\title{\sffamily Enhancing data locality of the conjugate gradient method for high-order matrix-free finite-element  implementations}

\author{ Martin Kronbichler\thanks{Technical University of Munich, Garching, Germany (\texttt{\{kronbichler,munch\}@lnm.mw.tum.de}).} \thanks{University of Augsburg, Augsburg, Germany (\texttt{martin.kronbichler@uni-a.de}).}%
  \and%
  Dmytro Sashko\thanks{The University of Queensland, Australia
    (\texttt{dmshko@gmail.com}).}%
  \and%
  Peter~Munch\footnotemark[1]~\thanks{Helmholtz-Zentrum Hereon, Geesthacht, Germany (\texttt{peter.muench@hzg.de}).}
}

\date{
  \begin{minipage}{0.8\textwidth}
  \small
  \noindent {\sf \textbf{Abstract.}}
This work investigates a variant of the conjugate gradient (CG) method
and embeds it into the context of high-order
finite-element schemes with fast matrix-free operator evaluation and cheap
preconditioners like the matrix diagonal. Relying on a
data-dependency analysis and appropriate enumeration of degrees of freedom, we interleave the vector
updates and inner products in a CG iteration with the matrix-vector
product with only minor organizational overhead. As a result, around
90\% of the vector entries of the three active vectors of the CG
method are transferred from slow RAM memory exactly once per iteration,
with all additional access hitting fast cache memory.
Node-level performance analyses and scaling studies on up to 147k cores show
that the CG method with the proposed performance optimizations is around two times faster than a standard CG solver as
well as optimized pipelined CG and $s$-step CG methods for large sizes that
exceed processor caches, and provides similar performance near the strong
scaling limit.\medskip
\end{minipage}
}

\maketitle

\input{tables.tex}

\section{Introduction}
\label{sec:intro}

The conjugate gradient (CG) method is one of the most popular algorithms
for the iterative solution of large sparse symmetric positive-definite linear systems arising
from discretization of partial differential equations. While it needs to be
combined with strong preconditioners such as multigrid when applied to
elliptic equations, the conjugate gradient method with simple preconditioners
like the matrix diagonal can be the most efficient choice for parabolic
partial differential {equations} with small to moderate time steps. For example
in computational fluid dynamics, many splitting schemes eventually lead to a
{positive definite} Helmholtz-like equation with a mass matrix and a diffusive operator scaled by
the time step and viscosity, see, e.g., \cite{Tufo99},
\citet[Sec.~6.5]{Deville02}, and \cite{Fehn18} for application in
incompressible flows as well as \cite{Demkowicz90} and \cite{Guermond21} for
compressible flows. Another important application is the projection with
consistent finite-element mass matrices, possibly including some
regularization through diffusion \citep{Kronbichler18multiphase}.

In the conjugate gradient method with simple preconditioners, the
matrix-vector product has traditionally been the most expensive
operation. With the increase in computing power through parallelism on the one
hand and algorithmic progress on the other hand, the matrix-vector product may
in fact be so cheap that attention must be turned to the other operations in
the CG method.

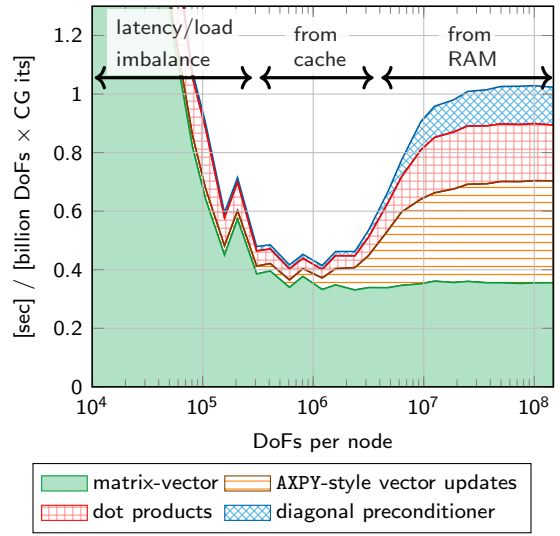
\begin{figure}
\centering
  \begin{tikzpicture}
    \begin{semilogxaxis}[
      width=0.9\columnwidth,
      height=0.78\columnwidth,
      xlabel={DoFs per node},
      ylabel={[sec] / [billion DoFs $\times$ CG its]},
      legend cell align={left},
      legend to name=legendCGBreakdown,
      legend columns = 2,
      grid,
      semithick,
      xmin=1e4,xmax=1.49e8,
      ymin=0,ymax=1.3,
      ytick={0,0.2,0.4,0.6,0.8,1,1.2},
      stack plots=y,
      area style,
      ]
      \addplot[color=gnuplot@green,fill=gnuplot@green!30] table[x index={1}, y expr={\thisrowno{3}/\thisrowno{1}*1e9}] {\tableMVDetailIntel}
      \closedcycle;
      \addplot[color=orange!50!black,fill=orange!30,pattern=horizontal lines,pattern color=orange] table[x index={1}, y expr={\thisrowno{4}/\thisrowno{1}*1e9}] {\tableMVDetailIntel}
      \closedcycle;
      \addplot[color=red,fill=red!14,pattern color=red!40,pattern=grid] table[x index={1}, y expr={\thisrowno{5}/\thisrowno{1}*1e9}] {\tableMVDetailIntel}
      \closedcycle;
      \addplot[color=gnuplot@darkblue,fill=gnuplot@lightblue!30,pattern color=gnuplot@lightblue,pattern=crosshatch] table[x index={1}, y expr={\thisrowno{6}/\thisrowno{1}*1e9}] {\tableMVDetailIntel}
      \closedcycle;
      \legend{{matrix-vector}, {\texttt{AXPY}-style vector updates}, {dot products},
        {diagonal preconditioner}};
    \end{semilogxaxis}
    \draw[<->,very thick,black] (0.03,4.1) -- node[above,outer sep=1pt,fill=white,opacity=0.85,text width=1.5cm]{\footnotesize\sffamily latency/load\\imbalance} (2.1,4.1);
    \draw[<->,very thick,black] (2.2,4.1) -- node[above,outer sep=1pt,fill=white,opacity=0.85,text width=0.6cm]{\footnotesize\sffamily from\\cache\\} (3.7,4.1);
    \draw[<->,very thick,black] (3.8,4.1) -- node[above,outer sep=1pt,fill=white,opacity=0.85,text width=0.5cm]{\footnotesize\sffamily from \\RAM\\} (6.05,4.1);
  \end{tikzpicture}
  \\
  \pgfplotslegendfromname{legendCGBreakdown}
  \caption{Breakdown of times per CG iteration in the CEED benchmark problem
    BP4~\citep{Fischer19} with finite elements of degree $p=5$
    on $2\times 24$ cores of Intel Xeon Platinum 8174. See
    Section~\ref{sec:cg_breakdown} below for a more detailed explanation of
    the steps.}
  \label{fig:ceed_breakdown}
\end{figure}

On large-scale parallel computers, the global reductions involved in the two
inner products in each CG iteration are generally seen as the main threat to
strong scaling, addressed by the development of lower-synchronization
variants, such as the pipelined conjugate gradient method
\citep{pipelined14,pipelined18} or $s$-step methods \citep{sstep1989}. These
alternatives rely on mathematical transformations of the basic CG algorithm
with redundant vector operations that break some dependencies. The $s$-step
method not only allows to combine global communication for several
CG iterations into one block, but also to schedule the communication of several matrix-vector
products together through matrix-power kernels.

The guiding theme of these recent contributions has been the reduction of the
\emph{communication latency}, see also \cite{Eller2019} for a broader overview
on large-scale methods. However, less attention has been paid to the
\emph{throughput of the memory hierarchy}, i.e., bandwidth {requirements} from and to main
memory (RAM). This can be the more severe performance limit in a number of applications, especially for
solvers that combine different algorithms in tight sequence.  One example is
incompressible fluid flow discretized with splitting methods, where the
pressure Poisson equation solved with multigrid sets the limit for strong
scaling, but the much larger symmetric positive definite system in the velocity
contributes with 50\% or more of the runtime \citep{Krank17,Fehn18}.  As an
illustration, Figure~\ref{fig:ceed_breakdown} shows the share of runtime of
different operations in a preconditioned conjugate gradient solver as a function of
problem size on a single compute node. While the matrix-vector product indeed
dominates the runtime for small sizes with less than 3 million degrees of
freedom, this is not the case for larger sizes relevant to those fluid
dynamics applications where \texttt{AXPY}-style vector updates, dot products
and the application of the preconditioner take up two thirds of the total run
time.

The aim of the present work is to design a solver with primary focus on the
memory bandwidth behavior of the CG algorithm in the context of high-order
finite-element methods implemented with matrix-free sum-factorization
algorithms \citep{Deville02}. The main novelty is {a set of techniques
that allow to interleave}
the vector updates and inner products in a CG
iteration with the matrix-vector product {for a specific pipelined-like CG formulation originally presented as Algorithm 2.2 in \cite{sstep1989}}. As a result, we are able to {perform the access to the three active vectors in inner products and vector updates of a complete CG iteration with a single load from RAM memory for}
around 90\% of the vector entries,
{serving all other accesses} from the fast cache memory on contemporary cache-based CPU
architectures. Our experiments show {similar performance as for}
pipelined and $s$-step methods near the strong scaling limit when all vector
entries are hit in the caches, but we reach a significantly higher throughput
when the vectors {spill out of the} caches. While not directly reducing the
minimum achievable wall time, our contribution allows to reach a predefined
throughput already on a smaller machine.

The {proposed techniques rely} on introspection of the matrix-vector product and
simple preconditioners. The idea of using the structure of the operations in
the CG iteration to increase performance is not new and can be traced back to
at least \cite{Eisenstat81}. However, the context of minimizing data movement
for high-order finite-element solvers within a single iteration appears to be
novel. These developments are necessary, because the wide stencils from
high-order finite-element methods as well as multi-component systems make
traditional optimizations such as matrix-power kernels and temporal wavefront
blocking \citep{malas2017multidimensional} in the context of $s$-step Krylov
methods ineffective.

The implementations used for the
present study are available as open-source software on
GitHub.\footnote{\url{https://github.com/kronbichler/mf\_data\_locality},
  retrieved on May 12, 2022.} They build on the general-purpose
finite-element library deal.II \citep{dealiicanonical} and have been verified
on supercomputer scale \citep{Arndt20}. The remainder of this contribution is
structured as follows. Section~\ref{sec:matvec} introduces the state of the
art of fast matrix-free operator evaluation for higher-order finite-element
discretizations. In Section~\ref{sec:cg}, the classical conjugate gradient
algorithm as well as pipelined and $s$-step variants are reviewed in terms of
the memory access. Section~\ref{sec:improved_cg} {discusses} a variant of CG that
avoids the two synchronization points of the conventional CG algorithm when
using cheap diagonal preconditioning, whereas Section~\ref{sec:interleave}
presents the ingredients necessary to {efficiently embed the vector
operations}
into the matrix-vector product. In Section~\ref{sec:results}, large-scale
computations are given to show the effectiveness of the method, before
Section~\ref{sec:conclusions} summarizes our results.

\section{Fast matrix-free operator evaluation}\label{sec:matvec}

{We consider a benchmark problem in the context of high-order finite element methods to investigate the benefits of the proposed techniques, in comparison to
well-studied optimized CG alternatives} from the literature.
{It involves} the vector-valued Poisson equation in a
$d=3$ dimensional domain $\Omega \subset \mathbb{R}^3$,
\begin{equation}\label{eq:poisson}
  -\nabla^2 \boldsymbol u = \boldsymbol f,
\end{equation}
with the vector field
$\boldsymbol u(\boldsymbol x) = (u_1(\boldsymbol x), u_2(\boldsymbol x),
u_3(\boldsymbol x))\in \left(H^{1}(\Omega)\right)^3$ and a forcing
$\boldsymbol f\in \left(L^2(\Omega)\right)^3$. On the domain boundary
$\partial \Omega$, Dirichlet boundary conditions
$\boldsymbol u = \boldsymbol g$ are set.

The finite-element discretization is derived from
the weak form of Equation~\eqref{eq:poisson}, restricted to a space of
polynomials on a mesh of elements $\Omega_e$ of the computational domain,
$e=1,\ldots,n_\text{cells}$. On a hexahedral element $\Omega_e$, the
solution interpolation is given by
\begin{equation}\label{eq:fem_ansatz}
  \boldsymbol u_h(\boldsymbol x)\big|_{\boldsymbol x \in \Omega_e} =
  \sum_{j=1}^{3(p+1)^3}\boldsymbol \phi_j(\hat{\boldsymbol x}(\boldsymbol x)) u_{e,j}.
\end{equation}
Here, $\myvector u_e = [u_{e,j}]_j$ 
denotes the vector of unknown coefficients
on $\Omega_e$ in an expansion with a polynomial basis
$\left\{\boldsymbol \phi_j, j=1,\ldots,3(p+1)^3\right\}$. The basis functions
are constructed as the tensor product of one-dimensional polynomials of degree
$p$ for each of the vector components. Collecting the functions defined on all
the elements and inserting the expansions as tentative solutions and test
functions into the weak form, we arrive at a matrix system
\begin{equation}\label{eq:linear_system}
  \myvector A \myvector u = \myvector b,
\end{equation}
with a sparse matrix $\myvector A \in \mathbb{R}^{n\times n}$, the
right-hand-side vector $\myvector b \in \mathbb{R}^n$ and the discrete
solution vector $\myvector u \in \mathbb{R}^n$. The number
$n\sim 3 n_\text{cells} p^3$ denotes the number of degrees of freedom (DoFs),
counting the unique free coefficients in the expansion. The solution of this
matrix system is the subject of the present study.

The relatively dense coupling of degrees of freedom in the matrix stencil
makes sparse matrix-vector products in iterative solvers inefficient for
higher-order finite elements with degree $p\geq 2$. Considerable speedups can
be obtained by replacing the sparse matrix-vector product by a matrix-free
evaluation of the action of the matrix on a vector. Whereas stencil-like
approaches are most beneficial for the lowest-order elements on structured
meshes \citep{Bauer2018}, the method of choice for hexahedral elements with
general deformed shapes and higher degrees is to compute the integrals
underlying the finite-element method on the fly
\citep{Deville02,Brown10,Kronbichler11,Fischer19}.  The matrix-vector product
is computed as a sum of cell-wise contributions,
\begin{equation}
  \label{eq:matrixfree_loop}
  \myvector v = \myvector A \myvector u =
  \sum_{e=1}^{n_\text{cells}} \myvector P_e^\mathsf{T} \myvector A_e \left( \myvector P_e \myvector u\right),
\end{equation}
where $\myvector A_e$ is the representation of the operator on element $\Omega_e$ and
$\myvector P_e$ denotes the local-to-global mapping of unknowns such that
$\myvector u_e = \myvector P_e \myvector u$ gives restriction of the global
solution vector $\myvector u$ to the element. The local operation
$\myvector A_e \myvector u_e$ is again implemented in a matrix-free fashion
without building the element stiffness matrix~$\myvector A_e$,
\begin{align}
  \label{eq:matrixfree_cell}
  &\left[\myvector A_e \myvector u_e\right]_{i} =
    \int_{\Omega_e} \left(\nabla \boldsymbol \phi_i\right)^\mathsf T \nabla \boldsymbol u_h \mathrm d \boldsymbol x =\\
  &\quad \sum_{q=1}^{n_\text{q}} \left(\hat{\nabla} \boldsymbol \phi_i\right)^\mathsf T
    \boldsymbol J_{e,q}^{-1} (w_q \det \boldsymbol J_{e,q})  \boldsymbol J_{e,q}^{-\mathsf T}
    \sum_{j=1}^{3(p+1)^3} \hat{\nabla} \boldsymbol \phi_j u_{e,j}\Big|_{\hat{\boldsymbol x}_q}.\nonumber
\end{align}
The integrals are approximated by numerical quadrature on $n_\text{q}$
points. In this work, we consider the BP4 benchmark problem proposed by
\cite{Fischer19}, which selects the tensor-product Gaussian quadrature formula
with $n_\text{q}=(p+2)^3$ points $\hat{\boldsymbol x}_q$ per cell and the
associated quadrature weight $w_q$. The integrals are transformed to reference
coordinates $\hat{\boldsymbol x}$ via a polynomial mapping
$\boldsymbol x(\hat{\boldsymbol x})$ and the derivatives in real space
$\nabla$ are transformed to derivatives in reference coordinates
$\hat{\nabla}$ by multiplication with the inverse and transpose of the
Jacobian
$\left[\boldsymbol J_e(\hat{\boldsymbol x})\right]_{ij} = \frac{\partial
  x_i}{\partial \hat{x}_j}$. The local result $\myvector A_e \myvector u_e$ is
obtained by evaluating Equation~\eqref{eq:matrixfree_cell} for all test
functions $\boldsymbol \phi_i$, $i=1,\ldots,3(p+1)^3$.

The efficiency of the matrix-free
algorithm~\eqref{eq:matrixfree_loop}--\eqref{eq:matrixfree_cell} crucially
depends on evaluating $\hat{\nabla} \boldsymbol u_h$ at the quadrature points
and the multiplication by the test function gradient
$\hat{\nabla} \boldsymbol \phi_i$ as well as the summation over quadrature
points, respectively. For tensor-product shape functions that are integrated
on a tensor-product quadrature formula, sum factorization allows to decompose
these two steps into a series of one-dimensional interpolations of total cost
$\mathcal O(p^{d+1})$ per element in $d$ dimensions (or $\mathcal O(p)$ per
unknown), compared to the naive evaluation cost of $\mathcal O(p^{2d})$.  The
sum-factorization approach has been developed in the context of the spectral
element method by \cite{Orszag80}, \cite{Patera84}, and \cite{Tufo99}, see
also the book by \cite{Deville02} as well as recent implementation and
vectorization studies by \cite{Kronbichler11,Kronbichler2019},
\cite{Swirydowicz19}, \cite{Fischer19}, \cite{Sun20}, \cite{Moxey20}, and
\cite{Kempf20}.

\subsection{Experimental setup}

Our experiments use the implementation of matrix-free operator evaluation in
the deal.II finite-element library \citep{dealii92,dealiicanonical}, described
in \cite{Kronbichler11,Kronbichler2019}. The main computational kernels are
fully vectorized across elements, i.e.,
operation~\eqref{eq:matrixfree_cell} {is evaluated} on several cells for the different SIMD
lanes, and use an even-odd decomposition \citep{Solomonoff92} in sum
factorization to further reduce the arithmetic cost. The solution vectors
store unique unknowns, which necessitates indirect addressing for the access
of elemental data, represented as a matrix $\myvector P_e$ in
Equation~\eqref{eq:matrixfree_loop}. Indirect addressing involves additional
instructions compared to duplicating unknowns shared by several cells as used,
e.g., in Nek5000 \citep{nek5000}, but avoids redundant storage and speeds up
the other parts of the solver. In our implementation, the indices describing
$\myvector P_e$ use a compressed format of $3^3$ four-byte integers, from
which all $3\times (p+1)^3$ indices are deduced on the fly. The meshes are
partitioned by space-filling curves according to \cite{Bangerth11}.

The code has been compiled with the GNU compiler g++, version 9.2, with
optimization flags \texttt{-O3 -march=native -funroll-loops}{, which is the compiler with the best performance among GNU, Intel and clang for our code}. The experiments
have been conducted within a pure MPI setting. To reduce the overhead due to
communication between processes within a single compute node, we perform the
exchange of ghost values manually via \texttt{memcpy} and MPI-3.0
shared-memory features \citep{munch2020hyper}, instead of relying on {plain}
\texttt{MPI\_Isend} and \texttt{MPI\_Irecv}.

Following the benchmark description by \cite{Fischer19}, the algorithms are
mainly compared in terms of the throughput, i.e., the number of degrees of
freedom processed per second (DoFs/s) for one matrix-vector product in this
section or one iteration of the conjugate gradient method in the subsequent
sections. The throughput is obtained by the ratio of the number of degrees of
freedom in the linear system and the measured runtime. The runtime is taken as
the minimum of two separate jobs with four experiments each in order to reduce
the noise caused by other concurrent jobs on the supercomputer. Apart from
isolated outliers, the arithmetic mean of those eight runs is within 2\% of
the reported minimum.

Unless noted otherwise, the numerical experiments are run on a
dual-socket Intel Xeon Platinum 8174 (Skylake) system of the
supercomputer SuperMUC-NG.\footnote{\url{https://top500.org/system/179566/},
  retrieved on January 4, 2021.} The CPU cores run at a fixed frequency of 2.3
GHz, which gives an arithmetic peak of 3.5 TFlop/s. The 96 GB of random-access
memory (RAM) are connected through 12 channels of DDR4-2666 with a theoretical
bandwidth of 256 GB/s and an achieved STREAM triad memory throughput of 205
GB/s.

\subsection{Identification of fast matrix-vector product}\label{sec:matvec:fast}

Contemporary implementations of matrix-free methods with sum factorization
often precompute and store the metric terms in
$\boldsymbol J_{e,q}^{-1} (w_q \det \boldsymbol J_{e,q}) \boldsymbol
J_{e,q}^{-\mathsf T}$ at each quadrature point and load them during operator
evaluation. The precomputed setup is applicable to deformed (curvilinear) cells
and to variable coefficients. As shown in \cite{KronbichlerLjungkvist19}, the
evaluation~\eqref{eq:matrixfree_loop}--\eqref{eq:matrixfree_cell} is then
memory-bound on modern hardware. For an implementation that aims to maximize
the throughput for cell integrals according to \cite{Kronbichler2019},
it might be more economic to evaluate the metric terms on the fly
as well. To identify a suitable method, we compare the following variants
regarding the terms representing the geometric factors:
\begin{itemize}
  \setlength{\itemsep}{0pt}
  \setlength{\parskip}{0pt}
  \setlength{\parsep}{0pt}
\item tri-quadratic geometry evaluated on the fly from $3^3=27$ points
  (`quadratic geomet. compute'), loading $27\times 3$ doubles per cell, giving
  a matrix-vector product with 395 Flops/DoF for $p=5$,
\item geometry evaluated on the fly from $(p+2)^3$ points at the position of
  the quadrature points (`isoparametric compute'), loading $3$ doubles per
  quadrature point, yielding 417 Flops/DoF for \mbox{$p=5$},
\item precompute and load the inverse Jacobian
  $\boldsymbol J^{-\mathsf 1}_{e,q}$ and the Jacobian determinant times
  quadrature weight (`inverse Jacobian load') at each quadrature point,
  loading $10$ doubles per quadrature point, yielding 316 Flops/DoF for \mbox{$p=5$},
\item precompute and load the final symmetric coefficient tensor,
  $\boldsymbol J_{e,q}^{-1} (w_q \det \boldsymbol J_{e,q}) \boldsymbol
  J_{e,q}^{-\mathsf T}$ (`final tensor load') at each quadrature point,
  loading $6$ doubles per quadrature point, yielding a matrix-vector product
  with 267 Flops/DoF for $p=5$, as done, e.g., in~\cite{Swirydowicz19,Fischer19}.
\end{itemize}

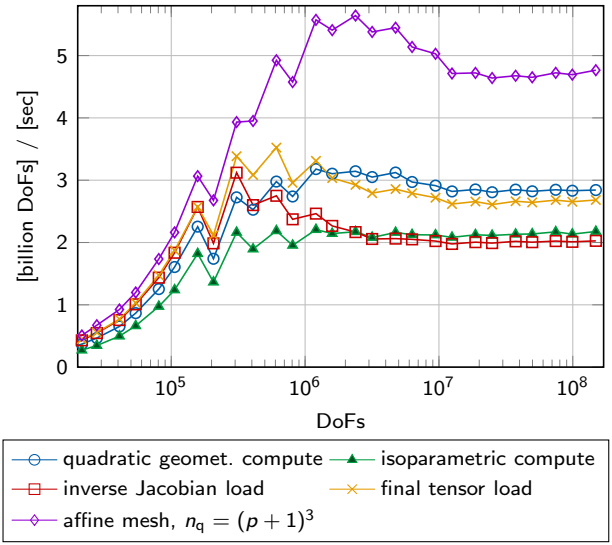
\begin{figure}
\centering
  \begin{tikzpicture}
    \begin{semilogxaxis}[
      width=\columnwidth,
      height=0.75\columnwidth,
      xlabel={DoFs},
      ylabel={[billion DoFs] / [sec]},
      legend cell align={left},
      cycle list name=colorGPL,
      legend to name=legendMVCompare,
      legend columns = 2,
      semithick,
      xmin=2e4,xmax=1.7e8,
      ymin=0,ymax=5.8,
      ytick={0,1,2,3,4,5},
      grid
      ]
      \addplot table[x index={1}, y expr={\thisrowno{1}*1e-9/(min(\thisrowno{7},\thisrowno{13})}] {\tableMVDetailIntel};
      \addlegendentry{quadratic geomet.~compute};
      \addplot[gnuplot@green,every mark/.append style={fill=gnuplot@green!50!black},mark=triangle*] table[x index={1}, y expr={\thisrowno{1}*1e-9/(min(\thisrowno{9},\thisrowno{15})}] {\tableMVDetailIntel};
      \addlegendentry{isoparametric compute};
      \addplot[gnuplot@red,every mark/.append style={fill=gnuplot@red!50!black},mark=square] table[x index={1}, y expr={\thisrowno{1}*1e-9/(min(\thisrowno{8},\thisrowno{14})}] {\tableMVDetailIntel};
      \addlegendentry{inverse Jacobian load};
      \addplot table[x index={1}, y expr={\thisrowno{1}*1e-9/(min(\thisrowno{10},\thisrowno{16})}] {\tableMVDetailIntel};
      \addlegendentry{final tensor load};
      \addplot table[x index={1}, y expr={\thisrowno{1}*1e-9/(min(\thisrowno{12},\thisrowno{18})}] {\tableMVDetailIntel};
      \addlegendentry{affine mesh, $n_\text q=(p+1)^3$};
    \end{semilogxaxis}
  \end{tikzpicture}
  \\
  \pgfplotslegendfromname{legendMVCompare}
  \caption{\sansmath Comparison of different implementations of matrix-free
    operator evaluation for polynomial degree $p=5$ on $2\times 24$ cores of
    Intel Xeon Platinum 8174.}
  \label{fig:mv_compare}
\end{figure}

Figure~\ref{fig:mv_compare} compares the computational throughput of these
variants on a single compute node. The operator evaluation reaches a maximum
for intermediate sizes of around $10^6$ DoFs when most data fits into
caches. As the problem size further increases, data must be fetched from main
memory, leading to a slowdown for the cases that are dominated by memory
access. {We note a slight zig-zag pattern in the reported
  throughput, which is caused by different costs of ghost exchange, which
  changes when the number of cells is divisible by 48 leading to cube-like
  subdomains (higher throughput) or by 64 leading to more irregular MPI
  subdomains (lower throughput).} Figure~\ref{fig:mv_compare} also presents
the throughput of the evaluation on an affine mesh with a constant inverse
Jacobian $\boldsymbol J^{-1}$ throughout the whole mesh and using
$n_q=(p+1)^d$ points of Gaussian quadrature, a case studied in detail in
\cite{Kronbichler11,Kronbichler2019}. This reduces the arithmetic cost to 206
Flops/DoF and the memory transfer to just the input and output vectors with
performance mainly limited by the vector access with indirect addressing.

For large sizes with $n>10^7$, the `load' variants are memory-limited at
slightly more than 200 GB/s, whereas the two `compute' variants involve a
memory transfer of 100 GB/s and 140 GB/s for vector sizes of 100 million DoFs,
all measured from hardware performance counters with the LIKWID tool
\citep{likwid10}.  Albeit slightly slower than the `load' variants for the
in-cache case with $n<10^6$, this study concentrates on the quadratic geometry
representation evaluated on the fly with polynomial degree $p=5$ for the
finite-element expansion~\eqref{eq:fem_ansatz}.  The representation of curved
geometries differs from the other three options in general, but we argue that
a tri-quadratic approximation is nonetheless suitable for many
applications. The bulk of a 3D geometry can often be well-represented in such
a way, leading to a significant reduction of the memory transfer and cache
pressure against the isoparametric high-order case. By contrast, a tri-linear
representation (with approximately 10\% higher throughput) might be
unacceptable in a whole region around strongly curved boundaries. It is conceivable to augment
the present strategy with a high-degree (isoparametric) geometry
representation of one element layer close to the boundary, without
significantly affecting the throughput.

From the throughput values listed in Figure~\ref{fig:mv_compare} and the
operation counts mentioned above, it can be deduced that the matrix-vector
product with quadratic geometry runs at 1.1 TFlop/s with 50 million DoFs and
at 1.3 TFlop/s with 1.2 million DoFs. While this is clearly below the
arithmetic peak of 3.5 TFlop/s, the value is high for this kind of algorithm;
the gap to the peak can be explained by the cost of the indirect addressing
into the vectors $\myvector u, \myvector v$, isolated additions and
multiplications that cannot be merged into fused multiply-add operations, the
throughput of caches, and, for the larger case, insufficient data prefetching
from RAM.

The throughput of 2.82 billion DoFs/s with 50 million DoFs for a matrix-free
operator evaluation ($p=5$, quadratic geometry computation) can be compared to
a sparse matrix-vector product: the lowest order $p=1$ can reach a throughput
of between 590 million DoFs/s (separate matrix entries for all three vector
components, perfect caching of vector entries) and 1.6 billion DoFs/s (same
matrix for all three vector components; only applicable for simple boundary
conditions), or between 50 and 147 million DoFs/s for the $p=5$ case. The
effect of high-order matrix-free algorithms being several times faster than
low-order matrix-based algorithms on a degree-of-freedom basis has been
examined in detail, e.g.,~in \cite{Kronbichler18}.

\section{Conjugate gradient algorithm}\label{sec:cg}

The high throughput of the matrix-free operator evaluation has important
implications for performance tuning of the CG iterative method
as the matrix-vector product might no longer be the dominant operation.
Despite using an accurate integration with $p+2$ points per direction, the
throughput shown in Figure~\ref{fig:mv_compare} is around a third of that of
simply copying one vector to the other, which achieves a throughput of 8.5
billion DoFs/s at 205 GB/s due to 24 bytes of access per unknown with 8 bytes
read, 8 bytes write, 8 bytes of read-for-ownership transfer \citep{Hager11} on
a dual-socket Intel Xeon 8174 machine.

\begin{algorithm}[t]
\begin{algorithmic}[1] 
  \State $ \myvector r_{0}  = \myvector{b} - \myvector{A} \myvector x_{0} $,
  $ \myvector z_{0}  = \myvector M^{-1} \myvector r_{0} $,
 $ \myvector p_{0} = \myvector z_{0} $, $ e_0 = \myvector r_0^ \mathsf T \myvector z_0 $
  \State $ k = 0 $
      \While {not converged}

      \State $ \myvector{ v}_{k} = \myvector{A} \myvector{p}_k$

      \State $ \alpha_k  = \frac{\displaystyle e_k}{\myvector{p}_{k}^{\mathsf T} \myvector{v}_{k}} $ \tikzmark{top0} \tikzmark{bot0}

      \State $ \myvector{x}_{k + 1} = \myvector{ x}_{k} + \alpha_k \myvector{ p}_k $  \tikzmark{top1}
      \State $ \myvector{r}_{k + 1} = \myvector{ r}_{k} - \alpha_k \myvector{ v}_{k} $ \tikzmark{bot1}

      \If {$ \sqrt{\gamma_{k+1}} = \norm{ \myvector{r}_{k + 1} } < \epsilon $}
      \State \textbf{break}
      \EndIf

      \State $ \myvector{z}_{k+1} = \myvector{M}^{-1} \myvector{r}_{k+1} $

      \State $ e_{k+1} = \myvector r_{k+1}^ \mathsf T \myvector z_{k+1} $\tikzmark{top2} \tikzmark{bot2}
      \State $ \beta_{k} = \frac{\displaystyle e_{k+1}} {\displaystyle e_k}$

      \State $ \myvector {p}_{k+1} =  \myvector{ z}_{k+1} + \beta_k \myvector{p}_{k} $\quad \tikzmark{right}\tikzmark{top3} \tikzmark{bot3}

      \State $ k = k + 1 $

      \EndWhile
\end{algorithmic}
\AddOneLineNote{top0}{bot0}{right}{\mbox{1st region: r:2}}
\AddLightNote{top1}{bot1}{right}{\mbox{\strut\ \ 2nd region: r:4/w:2}}
\AddOneLineNote{top2}{bot2}{right}{\mbox{3rd region: r:2}}
\AddOneLineNote{top3}{bot3}{right}{\mbox{\strut\ \ 4th region: r:2/w:1}}
\caption{ Preconditioned conjugate gradient method.}
\label{alg:precond_cg}
\end{algorithm}

For preconditioning, this work considers the case of a simple point Jacobi
preconditioner, i.e., the matrix diagonal. This preconditioner is
representative for problems including a strong mass-matrix contribution besides
the Laplacian~\eqref{eq:poisson}, as argued in \cite{Fischer19}. Since the same
coefficient is used for all 3 vector components of $\boldsymbol u$, only the
diagonal to a scalar Laplacian (computed with Gauss--Lobatto integration on
$p+1$ points) is stored and applied to all three components.

\subsection{Breakdown of runtime}\label{sec:cg_breakdown}

Figure~\ref{fig:ceed_breakdown} shows a breakdown of the runtime per
unknown {for one CG iteration}, plotted over the number of unknowns for the basic
CG variant presented in Algorithm~\ref{alg:precond_cg}. In this study, we
consider the termination by the unpreconditioned residual norm
$\|\myvector r_k\|$, which involves a third global reduction in each
iteration. Other variants exist, and the main performance characteristics
carry over similarly. The following kernels are considered:
\begin{itemize}
  \setlength{\itemsep}{0pt}
  \setlength{\parskip}{0pt}
  \setlength{\parsep}{0pt}
  \item the sparse matrix-vector product with matrix-free operator evaluation,
\item \texttt{AXPY}-like vector operations ( $\myvector{y} =  a\myvector{x} + \myvector{y}$ ),
\item dot product computations (including $l_2$ norm), and
\item the application of the diagonal preconditioner.
\end{itemize}
The \texttt{AXPY}-like vector operations and preconditioner application do not
involve any communication, the matrix-vector product communicates between
nearest neighbors in the mesh (e.g., 26 on a cube geometry with perfect
split), whereas the dot product involves a reduction among all participating
processes. The experiment of Figure~\ref{fig:ceed_breakdown} has been
conducted on a single compute node with 48 cores.

In the left part of the plot in Figure~\ref{fig:ceed_breakdown} with fewer
than $10^5$ DoFs, the load imbalance of the partitioning of the mesh elements
onto 48 processes as well as the latency of the communication between the
different cores on the node lead to an approximately constant runtime of
$6\times 10^{-5}$ seconds per iteration. This appears as a decrease of time
per unknown as the size increases in the figure. The \emph{latency}
limitations disappear for $n\sim 10^6$ DoFs, indicating a \emph{throughput}
limitation
instead with a plateau in timings per unknown. For very large sizes $n>10^7$,
the data set of the conjugate gradient exceeds the caches and most data needs
to be fetched from main memory (RAM).  Then, the vector operations start to
contribute significantly to the runtime, causing a severe slowdown compared to
intermediate sizes.

In order to understand the performance limitations of { the CG algorithm}, we take a closer
look at Algorithm~\ref{alg:precond_cg}. Treating the matrix-vector product and
the preconditioner as black boxes, there are four separate regions of vector
access in the form of dot products and \texttt{AXPY}-like vector
operations. Within each region, loop fusion leading to a single loop over the
entries of all vectors in the region may improve the locality of
reference.
Loop fusion can for example be used to compute the sum needed for the norm
$\|\myvector r_{k+1}\|$ already during the computation of $\myvector r_{k+1}$,
avoiding an extra vector load.

Between the regions, however, synchronization points prevent loop fusion and
all vector entries need to be touched before starting the next region.  For
instance, the computation of $\myvector{x}_{k+1}$ and $\myvector{r}_{k+1}$
depends on $\myvector{p}_k$, $\myvector{r}_k$, $\myvector{v}_k$,
$\myvector{x}_k$, and $\alpha_k$. The latter itself depends on $\myvector{p}_k$
and $\myvector{v}_{k}$ and requires a full vector sweep through
them.  If the size of the vectors $\myvector{p}_{k}$ and $\myvector{v}_{k}$
exceeds the capacity of a particular cache level
during the computation of the dot product for $\alpha_k$ and
the entries are already evicted from the cache in the
form of capacity misses, a second load  from the upper levels of the memory hierarchy is
inevitable. Similarly, during
the computation of $\myvector{p}_{k+1}$ in the forth region, the vector entries of
$\myvector{p}_{k}$ and $\myvector{z}_{k+1}$ would have to be loaded again,
despite being touched in the second region and inside the preconditioner,
respectively. Note that even with an ideal {cache replacement} strategy this
problem cannot be resolved for vectors considerably larger than the caches.

Summarizing the number of reads in each region of the conjugate gradient
algorithm, the preconditioned conjugate gradient algorithm requires 10 full
vector reads in each iteration besides the access for the matrix-vector
product and preconditioner, despite only 4 vectors participating in the
algorithm (assuming $\myvector v_k$ and $\myvector z_{k+1}$ use the same
memory). This number can be slightly reduced to 9 by moving the computation of
$\myvector{x}_{k+1}$ to the 4th region to reuse reads of $\myvector{p}_{k}$.

\subsection{Alternative CG methods}

For a simpler comparison, we now consider plain conjugate gradient algorithms
without preconditioner. The basic version (Algorithm~\ref{alg:precond_cg} with
$\myvector z_{k+1} = \myvector r_{k+1}$ and $\myvector{M}^{-1}=\myvector{I}$)
requires 9 full vector reads and 3 vector writes besides the access for the
matrix-vector product.

In the literature, a series of alternative flavors of the CG algorithm have
been developed with the goal to reduce the number of synchronization points,
primarily driven by \emph{latency} considerations. However, they naturally
also increase the possibility for loop fusion and might therefore {also} improve
the memory transfer \citep{rupp2016pipelined}. As a point of comparison of the algorithm structure, we
present Algorithm~\ref{alg:merged_pipelined} for the pipelined conjugate
gradient method and Algorithm~\ref{alg:merged_s_step} for the $s$-step
conjugate gradient methods, respectively.  To simplify the presentation,
hereafter we ignore the algorithms' initialization and focus on the structure
of the main iteration.

\begin{algorithm}[t]
\begin{algorithmic}[1] 
      \While {not converged} \tikzmark{right}

      \State $ \myvector{q}_k = \myvector{A} \myvector{w}_k $

	  \State $\beta_k = {\gamma_{k-1}} / \gamma_{k-2}$
	  \State $\alpha_k = \gamma_{k-1} / \left({a_{k-1}} - \beta_{k} \frac{\gamma_{k-1}}{\alpha_{k-1}}\right)$\quad\tikzmark{right4}

      \State $ \myvector{p}_k = \myvector{r}_k + \beta_k \myvector{p}_{k-1} $\qquad\tikzmark{scalstart}
      \State $ \myvector{x}_k = \myvector{x}_k + \alpha_k \myvector{p}_k $
      \State $ \myvector{s}_k = \myvector{w}_k + \beta_k \myvector{s}_{k-1} $
      \State $ \myvector{r}_k = \myvector{r}_{k-1} - \alpha_k \myvector{s}_k $
      \State $ \myvector{z}_k = \myvector{q}_k + \beta_k \myvector{z}_{k-1} $
      \State $ \myvector{w}_k = \myvector{w}_{k-1} - \alpha_k \myvector{z}_k $
      \State $ {\gamma_k} = \myvector{r}_k^\mathsf T \myvector{r}_k $
      \State $ {a_k} = \myvector{w}_k^\mathsf T \myvector{r}_k $ \tikzmark{scalend}

      \EndWhile
\end{algorithmic}
\AddLightNote{scalstart}{scalend}{right4}{r:7/w:6}
\caption{Pipelined conjugate gradient method.}
\label{alg:merged_pipelined}
\end{algorithm}
\begin{algorithm}[t]
\begin{algorithmic}[1] 
      \While {not converged} \tikzmark{right}

      \State $ \myvector{T}_k = \left[\myvector{r}_k,  \myvector{A} \myvector{r}_k, \dots, \myvector{A}^s \myvector{r}_k \right] $\quad\tikzmark{right4}

	  \State $\myvector{B}_k = - \myvector{W}_{k-1}^{-1}( \myvector{Q}_k^\mathsf T \myvector{P}_{k-1})$\smallskip\tikzmark{Bstart}\tikzmark{Bend}
	  \State $\myvector{P}_k = \myvector{R}_{k} + \myvector{P}_{k-1} \myvector{B}_{k}$\smallskip\tikzmark{Wstart}
	  \State $\myvector{W}_k = \myvector{Q}_{k}^\mathsf T \myvector{P}_{k}$\smallskip
	  \State $\myvector{g}_k = \myvector{P}_{k}^\mathsf T \myvector{r}_{k}$\smallskip\tikzmark{Wend}
	  \State $\myvector{a}_k = \myvector{W}_{k}^{-1} \myvector{g}_{k}$
	  \State $\myvector{x}_k = \myvector{x}_{k-1} + \myvector{P}_{k} \myvector{a}_{k}$\tikzmark{scalstart}\tikzmark{scalend}
	  \State $\myvector{r}_k = \myvector{b} - \myvector{A}_{k} \myvector{x}_{k}$ \tikzmark{Rstart}
	  \State $\gamma_k = \myvector{r}_k^\mathsf T \myvector{r}_k $\tikzmark{Rend}

      \EndWhile
\end{algorithmic}
\AddOneLineNote{scalstart}{scalend}{right4}{r:s+1/w:1}
\AddOneLineNote{Bstart}{Bend}{right4}{r:2s/w:0}
\AddLightNote{Wstart}{Wend}{right4}{r:2s+1/w:s}
\AddLightNote{Rstart}{Rend}{right4}{r:2/w:1}
\caption{ $s$-step conjugate gradient method with the aliases $\myvector{R}_k=\myvector{T}_k(:,1:s-1)$ and  $\myvector{Q}_k=\myvector{T}_k(:,2:s)$.}
\label{alg:merged_s_step}
\end{algorithm}

In the pipelined CG method~\citep{pipelined14}, the number of synchronization
points is reduced to one by introducing additional global auxiliary vectors.
Apart from the intended ability to overlap global
communication with the matrix-vector product, this also allows vector
operations to be concentrated in one vector access region.  A naive
implementation using a separate loop for each line of
Algorithm~\ref{alg:merged_pipelined} would yield a total of 15 vector reads
per CG iteration for the 7 participating vectors. Using loop fusion reduces
the number of reads to 7, the number of involved vectors. It is possible to
slightly reduce the memory transfer further by performing the update of
$\myvector{x}$ every other iteration (before and after the update of
$\myvector{p}$).

In contrast, $s$-step CG methods~\citep{sstep1989,naumov2016s} perform $s$ CG
iterations in a single phase, reducing the number of global reductions to 3
per phase, i.e., to $3/s$ per CG iteration. This is especially interesting
when the global reductions are the bottleneck of the CG algorithm. The global
reductions are aggregated by not working simply on vectors but on blocks of
$s$ vectors, e.g., $\myvector{P}_k$ instead of $\myvector{p}_k$ and
$\myvector{R}_k$ instead of $\myvector{r}_k$. Similarly, the scalar factor
$\alpha_k$ becomes a vector ($\myvector{a}_k\in \mathbb{R}^{s}$), $\beta_k$ a
matrix ($\myvector{B}_k \in \mathbb{R}^{s\times s}$), and dot products become
block dot products. The communication time of a block operation is similar to
that of a scalar one, since modern networks are latency-bound for global
reductions up to a few dozens of values.

In the literature, the operation
$[\myvector{A}\myvector{r}_k, ..., \myvector{A}^s\myvector{r}_k]$ is referred
to as a ``matrix-power kernel''.  It is typically considered to be uncritical
for performance, since it only comprises of $s$ point-to-point communication
steps in the worst case. For low-order methods, increasing the number of ghost
layers allows to use a single communication step per matrix-power-kernel
application~\citep{malas2017multidimensional}, which might be useful if the
latency is the limiting factor. Furthermore, it can also enable a higher
throughput of the matrix-vector product, since matrix and vector entries can
be held in caches. For the high-order (FEM) methods investigated here,
however, it does not pay off {according to preliminary investigations: The wide stencils lead to a much larger dependency region and quickly saturate caches. Already in the absence of communication, matrix-power-kernel applications
consisting of 3 matrix-vector products with the present high-order FEM for $p=5$ yield a lower throughput than performing three operator evaluations in sequence. Currently, we are not aware of more sophisticated implementations for this class of algorithms that could exploit this temporal locality. Furthermore, communication is negatively affected as additional ghost layers involve all unknowns on cells with a} high
surface-to-volume ratio~\citep{mehridehnavi2013communication}. As shown in
\cite{Kronbichler2019}, the cost of communicating all solution coefficients from
a single layer of elements is already substantial and leads to pronounced
slow-down of the matrix-vector product for $p>3$ in 3D.

In total, $s+1$ matrix-vector multiplications are performed per iteration and
four update regions can be identified with a total of $5s+4$ reads and $s+2$
writes per vector entry.  Finally, we would like to point out that the version
of the $s$-step CG method investigated in the following is numerically
unstable due to the loss of orthogonality of the monomial Krylov
subspace~\citep{naumov2016s}. However, as alternative formulations, which are
numerically more stable but involve additional steps, are structured
similarly, results obtained for this simple version are generally transferable
to other approaches.

{Similarly to the $s$-step CG methods, enlarged CG methods (ECG;
\cite{grigori2019scalable, lockhart2022performance}) also work on
blocks of vectors to accelerate convergence. The motivation for
the construction and the way to construct the blocks are somewhat
different, but the resulting high-level algorithms are similar
from the performance point of view to those of $s$-step CG.
Due to this similarity, we will not
consider ECG in the remainder of this work.}

\section{Minimize data access in standard CG}\label{sec:improved_cg}

{Inspired by the increased chances to fuse loops over vectors in the pipelined
and $s$-step conjugate gradient methods, we now study
a version of CG that has been introduced by \cite[Algorithm 2.2]{sstep1989} and served
as a starting point for the derivation of pipelining methods.
However, in the present work,
we do not further modify the algorithm by \cite{sstep1989} and instead
aim to reduce the main memory
transfer without introducing additional auxiliary vectors, that inherently increase the memory access.}

{{We start our derivation by noting that} the number of synchronization barriers identified in Algorithm~\ref{alg:precond_cg}
 can be reduced by using redundant computations of partial sums, which is possible in the case the preconditioner is cheap.

\subsection{No preconditioner}

We first consider the case of identity preconditioning
($\myvector{z}_k=\myvector{r}_k$ and $\myvector{M}^{-1} = \myvector{I}$) and aim to perform the computation of
contributions to
$\beta_{k}$ before finalizing the computation of $\alpha_k$ and
$\myvector{r}_{k+1}$. We therefore expand
$\myvector{r}_{k+1}^\mathsf T \myvector{r}_{k+1}$ into
\begin{equation}\label{eq:beta_update}
\begin{aligned}
\myvector{r}_{k+1}^\mathsf T \myvector{r}_{k+1} &=  (\myvector{r}_{k} - \alpha_k \myvector{v}_{k})^\mathsf T (\myvector{r}_{k} -
  \alpha_k \myvector{v}_{k}) \\
  \quad &= \myvector{r}_{k}^\mathsf T \myvector{r}_{k} -
  2 \alpha_k \myvector{r}_{k}^\mathsf T \myvector{ v}_{k} +
  \alpha_k^2 \myvector{v}_{k}^\mathsf T \myvector{v}_{k}.
\end{aligned}
\end{equation}
By computing the three sums for the inner products
$\myvector{r}_{k}^\mathsf T \myvector{r}_{k}$, $\myvector{r}_{k}^\mathsf T \myvector{ v}_{k}$,
$\myvector{v}_{k}^\mathsf T \myvector{v}_{k}$, the ingredients for
$\beta_{k}$ can be scheduled in parallel to the inner product
$\myvector{p}_{k}^\mathsf T \myvector{v}_{k}$ needed by $\alpha_k$, as shown in
Algorithm~\ref{alg:merged_cg}. {Note that
$\gamma_{k}=\myvector r_k^\mathsf T \myvector r_k$ is computed explicitly rather than defined} recursively from the
previous iteration {in order to avoid detrimental influence of roundoff
errors \citep{sstep1989,saad1985practical}}. While this {scheme} adds an additional read to
$\myvector {r}_k$ during the summation compared to the computation of
$\myvector{v}_{k}^\mathsf T \myvector{p}_{k}$ alone, this is compensated
by computing $\myvector{r}_{k+1}$ at the same time as using the respective entry for
$\myvector p_{k+1}$. In addition, the fused scheduling uses $\myvector p_{k}$ for both
$\myvector x_{k+1}$ and $\myvector p_{k+1}$. In the end,
number of vector access regions is reduced to 2,
one before (``pre'') and one after (``post'') the matrix-vector product.

\begin{algorithm}[t]
\begin{algorithmic}[1] 
      \State $k=0$,  $\alpha_0 = \beta_0 =0$, $\myvector{r}_{0} = \myvector{b} - \myvector{A} \myvector{x}_{0}$, $\myvector{p}_{0} = \myvector{v}_{0} = \myvector{0}$
      \While {not converged} \tikzmark{right}
      \State $k = k + 1$
      \If {$k>1$ odd} \tikzmark{vecstart}
      \State $ \myvector{x}_{k} = \myvector{ x}_{k-2} + \alpha_{k-1} \myvector{ p}_{k-1}$
      \State \phantom{abcd} $+ \frac{\alpha_{k-2}}{\beta_{k-2}}\left(\myvector{p}_{k-1} -  \myvector{r}_{k-1} \right)$\ \tikzmark{right3}
      \EndIf
      \State $ \myvector{r}_{k} = \myvector{r}_{k-1} - \alpha_{k-1} \myvector{ v}_{k-1} $
      \State $ \myvector{p}_{k} = \myvector{r}_{k} + \beta_{k-1} \myvector{ p}_{k-1} $ \tikzmark{vecend}
    \State $ \myvector{v}_k = \myvector{A} \myvector{p}_k $
      \State $ a_k = \myvector{p}_k^\mathsf T \myvector{v}_k  $ \tikzmark{scalstart}
      \State $ \gamma_k = \myvector{r}_k^\mathsf T \myvector{r}_k $
      \State $ c_k = \myvector{r}_k^\mathsf T \myvector{v}_k $
      \State $ d_k = \myvector{v}_k^\mathsf T \myvector{v}_k $\tikzmark{scalend}
      \State $ \alpha_k = \frac{\displaystyle \gamma_k}{\displaystyle a_k}$
      \State $ \gamma_{k+1} = \gamma_k-2\alpha_kc_k + \alpha^2_k d_k$
      \If  { $\sqrt{\gamma_{k+1}} < \epsilon$}
	    \If {$k$ odd}
        \State $ \myvector{x}_{k+1} = \myvector{x}_k + \alpha_k \myvector{p}_k $
        \Else
        \State $\myvector{x}_{k+1} = \myvector{ x}_{k-1} + \alpha_{k} \myvector{ p}_{k} + \frac{\alpha_{k-1}}{\beta_{k-1}}\left(\myvector{p}_{k} - \myvector{r}_{k} \right)$
        \EndIf
        \State \textbf{break}
      \EndIf
      \State $ \beta_{k} = \frac{\displaystyle \gamma_{k+1}}{\displaystyle \gamma_k} $

      \EndWhile
\end{algorithmic}
\AddLightNote{vecstart}{vecend}{right3}{\small\textbf{``pre''} region: \\r:3.5/w:2.5}
\AddLightNote{scalstart}{scalend}{right3}{\small\textbf{``post''} region:\\r:3/w:0}
\caption{ Conjugate gradient method with merged vector operations.}
\label{alg:merged_cg}
\end{algorithm}

It is also possible to perform the updates to $\myvector{x}_{k+1}$ only every other
iteration, reusing the content of the vector $\myvector{p}_{k-1}$ and
$\myvector{r}_{k-1}$ before they get updated. All together, the number of
vector reads is reduced from 9 in the basic CG iteration to 6.5 in this
improved variant.

{\cite{rupp2016pipelined} identified possibilities for additional performance
optimizations by the three phases ``pre'', ``matrix-vector product'', and ``post''. Specifically, that contribution proposed to merge the matrix-vector product
with the ``post'' region on the GPU for matrix-vector products through
sparse matrix representations in order to reduce the number of kernel calls. Building upon
this idea, we aim to merge \textit{both} regions with the matrix-vector product,
which on the one hand allows to reduce the memory transfer on the CPU,
but is also more involved in the context of matrix-free FEM.}

\subsection{Diagonal preconditioner}

The ideas of the previous subsection can be extended to the case of a
preconditioner. Under the assumption that the preconditioner is cheap and that
there are no long-range dependencies introduced to the computation of
$\myvector z_{k+1} = \myvector {M}^{-1} \myvector r_{k+1}$, it is more
economic to apply the preconditioner several times.

Following Equation~\eqref{eq:beta_update}, we decompose the computation of the
numerator for $\beta_{k}$ into several inner products that do not depend on
$\alpha_k$,
\begin{equation}\label{eq:beta}
\begin{aligned}
  \beta_{k} &= \frac{\myvector{z}_{k + 1}^\mathsf T \myvector{r}_{k+1}} {\myvector{z}_{k}^\mathsf T \myvector{r}_{k}} =   \frac{
    (\myvector{M}^{-1} \myvector{r}_{k+1})^\mathsf T \myvector{r}_{k+1}}{\myvector{z}_k^\mathsf T \myvector{r}_{k}} = \\
  &=\frac{\myvector{r}_{k}^\mathsf T \myvector {M}^{-1} \myvector{r}_{k} -
    2 \alpha_{k}  \myvector{r}_{k}^\mathsf T \myvector {M}^{-1} \myvector{v}_{k} +
    \alpha_k^{2} \myvector{ v}_{k} ^\mathsf T \myvector{M}^{-1}
    \myvector{v}_{k}}{\myvector{r}_{k}^\mathsf T \myvector {M}^{-1} \myvector{r}_{k}}.
\end{aligned}
\end{equation}
Thus, $\beta_{k}$ can be obtained only based on the value of $\myvector{r}_{k}$
and $\myvector{v}_{k}$ from the beginning of the iteration, prior to the
update of the vectors $\myvector{x}_{k+1}$, $\myvector{r}_{k+1}$,
$\myvector{z}_{k+1}$.

Similarly, the value of $\gamma_{k+1}=\norm{\myvector{r}_{k+1}}^2$ for the
convergence criterion can be computed in parallel to the reduction for
$\alpha_k$, using the expansion
\begin{align}\label{eq:gamma}
  \gamma_{k+1} &=  \myvector{r}_{k}^{\mathsf T} \myvector{r}_{k} - 2 \alpha_{k}  \myvector{r}_{k}^{\mathsf T}
                 \myvector{v}_{k}  + \alpha_k^{2} \myvector{v}_{k}^{\mathsf T} \myvector{v}_{k}.
\end{align}
Therefore, given the residual $\myvector{r}_k$ and the result of the
matrix-vector product $\myvector{v}_k$, all scalars of the current conjugate
gradient iteration can be computed using one reduction region. The vector
$\myvector{z}_{k+1}$ is no longer stored explicitly, since we assume that
the application of the preconditioner is cheaper than the read and write of
$\myvector{z}_{k+1}$. As a result of this restructuring, all vector updates
can be clustered in a single region. Simplifying the notation and combining
the expressions above, we obtain Algorithm~\ref{alg:merged_precond_cg}.

\begin{algorithm}[t]
\begin{algorithmic}[1] 
      \State $k=0$,  $\alpha_0 = \beta_0 =0$, $\myvector{r}_{0} = \myvector{b} -\myvector{A} \myvector{x}_{0}$, $\myvector{p}_{0} = \myvector{v}_{0} = \myvector{0}$

      \While {not converged} \tikzmark{right}
      \State $k = k + 1$

      \If {$k>1$ odd} \tikzmark{vecstart}
      \State $ \myvector{x}_{k} = \myvector{ x}_{k-2} + \alpha_{k-1} \myvector{ p}_{k-1}$
      \State $\strut\ + \frac{\alpha_{k-2}}{\beta_{k-2}}\left(\myvector{p}_{k-1} - \myvector M^{-1} \myvector{r}_{k-1} \right)$\ \tikzmark{right3}
      \EndIf
      \State $ \myvector{r}_{k} = \myvector{r}_{k-1} - \alpha_{k-1} \myvector{ v}_{k-1} $
      \State $ \myvector{p}_{k} = \myvector{ M}^{-1} \myvector{r}_{k} + \beta_{k-1} \myvector{ p}_{k-1} $ \tikzmark{vecend}
    \State $ \myvector{v}_k = \myvector{A} \myvector{p}_k $
      \State $ \gamma_k = \myvector{r}_k^\mathsf T \myvector{r}_k $ \tikzmark{scalstart}
      \State $ a_k = \myvector{p}_k^\mathsf T \myvector{v}_k  $
      \State $ b_k = \myvector{r}_k^\mathsf T \myvector{v}_k $
      \State $ c_k = \myvector{v}_k^\mathsf T \myvector{v}_k $
      \State $ d_k = \myvector{r}_k^\mathsf T \myvector{M}^{-1} \myvector{r}_k $
      \State $ e_k = \myvector{r}_k^\mathsf T \myvector{M}^{-1} \myvector{v}_k $
      \State $ f_k = \myvector{v}_k^\mathsf T \myvector{M}^{-1} \myvector{v}_k $ \tikzmark{scalend}
      \State $ \alpha_k = \frac{ d_k}{ a_k}$
      \If  { $\sqrt{\gamma_k-2\alpha_kb_k + \alpha^2_k c_k} < \epsilon $}
	    \If {$k$ odd}
        \State $ \myvector{x}_{k+1} = \myvector{x}_k + \alpha_k \myvector{p}_k $
        \Else
        \State {\small $\myvector{x}_{k+1} = \myvector{ x}_{k-1} + \alpha_{k} \myvector{ p}_{k} + \frac{\alpha_{k-1}}{\beta_{k-1}}\left(\myvector{p}_{k} - \myvector M^{-1} \myvector{r}_{k} \right)$}
        \EndIf
        \State \textbf{break}
      \EndIf
      \State $ \beta_{k} = \frac{\displaystyle d_k - 2\alpha_k e_k + \alpha_k^2 f_k}{\displaystyle d_k} $

      \EndWhile
\end{algorithmic}
\AddLightNote{vecstart}{vecend}{right3}{\small\textbf{``pre''} region:\\r:3.8$\overline{3}$/w:2.5}
\AddLightNote{scalstart}{scalend}{right3}{\small\textbf{``post''} region:\\r:3.$\overline{3}$/w:0}
\caption{ Preconditioned conjugate gradient method with merged vector operations.}
\label{alg:merged_precond_cg}
\end{algorithm}

The reformulated algorithm results in two vector access regions,
with loop fusion applicable within each region.  As in
Algorithm~\ref{alg:merged_cg}, the $\myvector{x}_k$ update can be delayed and
performed only every other iteration. The first fused loop region now consists
of 3.5 full vector loads per iteration, plus the load of the preconditioner
that---under the assumption that a single diagonal is used for each component
of the block PDE system~\eqref{eq:poisson}---consists of $\frac{1}{3}$
doubles per vector entry. The number of stores is one for $\myvector r_k$ and
one for $\myvector p_k$ as well as one for $\myvector x_k$ every second
iteration. The number of {vector} loads in the second region is equal to 3 plus $\frac 13$ for the diagonal preconditioner. The present
reformulation results in seven global reductions, which can be computed by
summations local to each MPI process and a single \texttt{MPI\_Allreduce}
carrying 7 variables. Once the seven scalars are available, the coefficients
$\alpha_{k}$ and $\beta_{k}$ can be computed locally. The presented
reformulation also enables a fusion into the matrix-vector product, as
discussed in the next section.

The proposed algorithm relies on three properties of the
preconditioner $\myvector M^{-1}$:
\begin{itemize}
  \setlength{\itemsep}{0pt}
  \setlength{\parskip}{0pt}
  \setlength{\parsep}{0pt}
\item The preconditioner, which is applied twice in the first vector access region and
  twice in the second vector access region, { is assumed to be cheap to apply, with arithmetic costs}
  hidden behind the memory transfer of the involved vectors.
\item {We assume that} there are no long-range dependencies in the preconditioner, allowing to
  reuse the respective entries of $\myvector r_k$ and $\myvector v_k$ from
  caches or registers {when $\myvector M^{-1}\myvector r_k$ and $\myvector M^{-1}\myvector v_k$ are computed}.
\item The memory access induced by the preconditioner is {assumed to be less} expensive than the
  aggregated store and load of $\myvector z_{k+1}$ in
  Algorithm~\ref{alg:precond_cg}.
\end{itemize}
A diagonal preconditioner obviously fulfills these properties, whereas, on
the other extreme, a multigrid V-cycle would violate all three requirements. {Clearly, it needs to be examined for each preconditioner whether it fits into this scheme on a case-by-case basis, with preconditioners with more global action requiring a separate storage step to get $\myvector M^{-1} \myvector r_{k+1}$ before the reductions for $\beta_k$.}

\section{Combining vector updates with matrix-vector product}\label{sec:interleave}

In the previous section, the matrix-vector product has been treated as a black
box.  In order to further improve the data reuse between the vector
access regions of Algorithms~\ref{alg:merged_cg} and
\ref{alg:merged_precond_cg}, we propose to embed the vector updates and dot
products into the matrix-free operator evaluation, which allows to {re-use hits of the entries of $\myvector{p}$, $\myvector{r}$, $\myvector{v}$ in caches during the ``post'' stages, leading to a single memory read of} the
vectors $\myvector{p}$, $\myvector{r}$, $\myvector{v}$, and $\myvector{x}$ in the ideal case (3.8$\overline{3}$ doubles per unknown).

This is realized by performing the operations identified in the previous
section on subranges of the vectors while looping over cells according to
Equation~\eqref{eq:matrixfree_loop} to exploit temporal locality.  In order to
produce a valid algorithm, the data dependencies during the matrix-vector
product need to be identified and translated into subranges, as detailed in
the following three subsections.  Note that this approach is more
{involved} than previously proposed algorithms that fuse vector operations
following the matrix-vector product into the loop over { unknowns in sparse matrix-vector products \citep{rupp2016pipelined} or over} cells for
discontinuous Galerkin schemes, e.g., in
\cite{Kronbichler18enviroinfo}, \cite{Charrier19} and \cite{munch2020hyper}.

\subsection{Data dependencies in matrix-free loops}\label{sec:dda}

On a high level, the matrix-vector multiplication depends on the source vector
$\myvector{u}$ and produces the destination vector $\myvector{v}$. However,
due to the cell-wise nature of our matrix-free algorithm, which
\begin{itemize}
  \setlength{\itemsep}{0pt}
  \setlength{\parskip}{0pt}
  \setlength{\parsep}{0pt}
\item runs through each cell of the mesh,
\item reads all unknowns $\myvector{u}_e = \myvector{P}_e \myvector{u}$
  attached to a cell, which can be shared with other cells for continuous
  finite elements, and
\item accumulates integral contributions in the same
way ($\myvector{P}_e^{\mathsf T}\myvector{v}_e$),
\end{itemize}
we can refine the dependency statement for each entry of the source and the
destination vector: the entry $u_i$, $i\in \{1,\ldots,n\}$, is only needed
once the first cell reads its value. Conversely, entry $v_i$ is available as
soon as the last cell has added its contribution. From an implementation point
of view, this means that we can postpone the update of $u_i$ until its first
usage and use the value of $v_i$ for the dot product as soon as its value has
been finalized. In the following, we are going to refer to operations
happening before the first read access to $\myvector u$ but still within the
matrix-free loop---in line with the region names in Algorithms~\ref{alg:merged_cg} and \ref{alg:merged_precond_cg}---as a ``pre'' operation and to operations happening after the
last write access to $\myvector v$ as a ``post'' operation.

\begin{figure}
  \centering
  \begin{tikzpicture}[scale=1]
    \fill[black!10] (3,2) rectangle (4,3);
    \foreach \y in {0,1,2,3,4}
      \draw (0,\y) -- (5,\y);
    \foreach \x in {0,1,2,3,4,5}
    \draw (\x,0) -- (\x,4);
    \foreach \x in {0,0.3,0.7,1,1.3,1.7,2,2.3,2.7,3,3.3,3.7,4,4.3,4.7,5}
    \foreach \y in {0,0.3,0.7,1,1.3,1.7}
    {
      \draw [thin,color=black!50] (\x-0.05,\y-0.05) -- (\x+0.05,\y+0.05);
      \draw [thin,color=black!50] (\x-0.05,\y+0.05) -- (\x+0.05,\y-0.05);
    }
    \foreach \x in {0,0.3,0.7,1,1.3,1.7,2,2.3,2.7}
    \foreach \y in {2,2.3,2.7}
    {
      \draw [thin,color=black!50] (\x-0.05,\y-0.05) -- (\x+0.05,\y+0.05);
      \draw [thin,color=black!50] (\x-0.05,\y+0.05) -- (\x+0.05,\y-0.05);
    }
    \foreach \x in {3,3.3,3.7}
      \foreach \y in {2,2.3,2.7}
      {
      \draw [thin] (\x-0.05,\y-0.05) -- (\x+0.05,\y+0.05);
      \draw [thin] (\x-0.05,\y+0.05) -- (\x+0.05,\y-0.05);
      }
    \foreach \x in {3.3,3.7,4}
    \foreach \y in {2.3,2.7,3}
    {
      \draw[thin] (\x,\y) circle (0.06);
    }
    \foreach \x in {0,0.3,0.7,1,1.3,1.7,2,2.3,2.7,3}
      \foreach \y in {3}
        \draw[thin] (\x-0.05,\y-0.05) -- (\x+0.05,\y-0.05) -- (\x+0.05,\y+0.05) -- (\x-0.05,\y+0.05) -- (\x-0.05,\y-0.05);
    \foreach \x in {4,4.3,4.7,5}
      \foreach \y in {2}
        \draw[thin] (\x-0.05,\y-0.05) -- (\x+0.05,\y-0.05) -- (\x+0.05,\y+0.05) -- (\x-0.05,\y+0.05) -- (\x-0.05,\y-0.05);
    \foreach \x in {0,0.3,0.7,1,1.3,1.7,2,2.3,2.7,3,3.3,3.7,4,4.3,4.7,5}
      \foreach \y in {3.3,3.7,4}
        \fill[black!20] (\x,\y) circle (0.06);
    \foreach \x in {4.3,4.7,5}
      \foreach \y in {2.3,2.7,3}
        \fill[black!20] (\x,\y) circle (0.06);
  \end{tikzpicture}
  \caption{Illustration of data dependencies in matrix-free operator
    evaluation with degree $p=3$ for a lexicographic loop through the cells
    starting from the bottom left. Each symbol represents an unknown. Gray
    crosses denote unknowns where the result of the operator evaluation is
    complete before the highlighted cell. The $3\times 3$ unknowns marked with
    black crosses get the final contribution from the highlighted cell and can
    schedule the ``post'' operation afterwards together with the unknowns
    marked with gray crosses. $3\times 3$ circles indicate unknowns that have
    their first access on the highlighted cell, thus necessitating to be
    preceded by the ``pre'' operation.  Black squares denote unknowns with
    pending integrals, i.e., the ``pre'' operation has already been done, but the
    ``post'' operation is not yet possible. Gray disks illustrate unknowns not
    yet processed.}
  \label{fig:dependencies_matvec}
\end{figure}
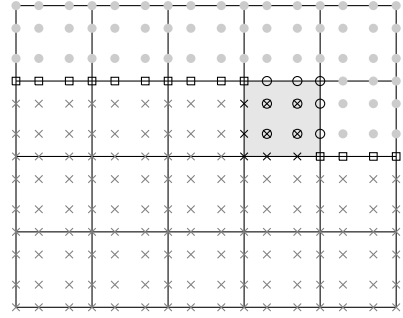

Figure~\ref{fig:dependencies_matvec} visualizes the data dependencies in a matrix-free
operator evaluation as well as its interplay with ``pre'' and ``post'' operations.
In an MPI-parallel context, the ghost exchange adds additional constraints
\citep[Algorithm 2.1]{Kronbichler11}. More precisely, all unknowns
owned by a process in the vector $\myvector u$ that need to be sent to remote
processes have to perform the ``pre'' operation before the ghost exchange is
initiated. Furthermore, the part of integrals accumulated on remote processes
need to be first sent to  owner of the respective entry in the
vector $\myvector v$ before the ``post'' operation can be scheduled on those
unknowns. It should, however, be noted that both communication
steps can be overlapped with computations on inner cells.

{We conclude this subsection by discussing the major differences to matrix-based
implementations. The popular compressed row storage and, similarly,
other sparse-matrix formats update an entry in the destination vector only once by
applying the whole row of the matrix. In such a context, it is obvious
when values in the destination vector are available and it is straightforward
to determine
when to schedule
the ``post'' operation during the matrix-vector product in a merged
way, as was exploited by \cite{rupp2016pipelined}. However, this relation
is not given for the dependency region of the source vector, allowing to
embed the ``pre'' operation closer to the user of vector entries only based on a dependency analysis similar to the one
proposed here.}

\subsection{Batching work from several cells}

Tracking the state of each individual vector entry $u_i, v_i$ for scheduling the ``pre'' and
``post'' operations would lead to excessive overhead and inhibit loop
optimizations, such as vectorization and unrolling of the vector operations in
CG. Therefore, the ``pre'' and ``post'' operations are tracked on ranges of
vector entries. The length of the range is given in multiples of 64, a
heuristic value that permits full vectorization with typical SIMD lengths
today, except for a single spot at the end of the vectors.

The length of the ranges is crucially influenced by the number of vector
entries processed by the matrix-free integrals in between. The intent is to
reach a significant share of overlap between the ``pre'' and ``post'' ranges,
enabling to reuse data read during the ``pre'' operations even during the
``post'' operations from the fast cache memory.  The range lookup and
the callback into matrix-free integration functions come with some overhead in
our implementation, which is especially noticeable for low and intermediate
polynomial degrees with small work per cell.  We therefore schedule the
``pre'' and ``post'' operations not around every individual cell, but around
\emph{batches of cells}. The size of the batches is selected as
\begin{equation}\label{eq:batch_size}
  n_\text{batch} = \max\left(\left\lfloor\frac{1024}{3(p+1)^3}\right\rfloor,
    2\right) n_\text{simd lanes}.
\end{equation}
The first expression inside the maximum operation
ensures that more cells are grouped together for lower polynomial degrees, by
dividing by the number of unknowns on each cell.
The resulting number of cells is multiplied by the number of SIMD lanes in the
instruction set in order to employ vectorization across elements
\citep{Kronbichler11}.
For higher degrees
($p \ge 4$), at least two SIMD groups of cells are used.

Depending on the number of SIMD lanes, the number of vectors accessed in the
``pre'' and ``post'' stages, as well as the additional data access for the
matrix-free integrals, the criterion given by Equation~\eqref{eq:batch_size}
leads to a few thousands to tens of thousands of double-precision values
corresponding to up to few hundreds of kB of data. This fits well within
modern level-2 or level-3 caches, which is why no additional tuning has been
performed.

{\textit{ We have integrated the proposed algorithm into deal.II
  \citep{dealii92}. It allows users to perform a ``pre'' and ``post''
  operation during any matrix-free loop by providing---additionally to
  the cell operation---appropriate anonymous functions in the style of:}
  \begin{align*}
  vmult(dst, src):=loop(dst, src, op\_cell, op\_pre, op\_post)
  \end{align*}
\textit{Since the operation $op\_cell$, which contains the specifics of
a the considered PDE/physics, is interchangeable, our approach is modular and
the proposed algorithms are simply applicable to other PDEs---not just those
considered in this publication. For CG, we provide of-the-shelf implementations of $op\_pre$ and $op\_post$.}}

\subsection{Numbering of unknowns}

The second ingredient is to minimize the number of ranges with a high number of cell batches
between the first and last access in the matrix-free loop. As seen from
Figure~\ref{fig:dependencies_matvec}, unknowns located on shared vertices,
edges, and faces all have the potential to reach over long distances. This
effect is exacerbated when working on blocks of 64 unknowns, because a single entry
out of 64 can lead to a delay of the ``post'' operation. It is therefore
crucial to develop a suitable cell traversal and numbering of unknowns. The
cell traversal should aim for a high volume-to-surface ratio of the cell
batches, because all unknowns located inside the cell batch have an optimal
pre-post distance. In this work, a Morton space-filling curve is used for the
partitioning of elements among the processes \citep{Burstedde11,Bangerth11}
and for the process-local mesh traversal.

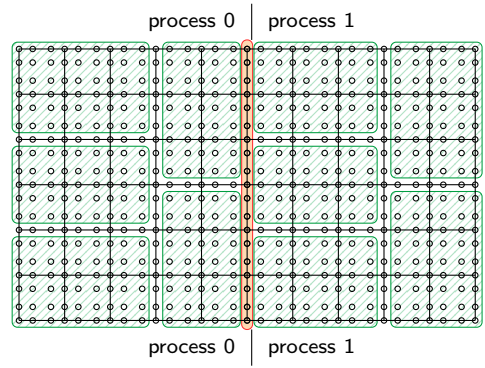
\begin{figure}
  \centering
  \begin{tikzpicture}[scale=0.6]
    \draw[red!80,fill=orange!30,rounded corners=2pt] (4.88,-0.2) rectangle (5.12,6.2);
    \draw[gnuplot@green,pattern color=gnuplot@green!30,rounded corners=2pt,pattern=north east lines,] (-0.15,-0.15) rectangle (2.85,1.85);
    \draw[gnuplot@green,pattern color=gnuplot@green!30,rounded corners=2pt,pattern=north east lines,] (-0.15,2.15) rectangle (2.85,3.85);
    \draw[gnuplot@green,pattern color=gnuplot@green!30,rounded corners=2pt,pattern=north east lines,] (-0.15,4.15) rectangle (2.85,6.15);
    \draw[gnuplot@green,pattern color=gnuplot@green!30,rounded corners=2pt,pattern=north east lines,] (3.15,-0.15) rectangle (4.85,2.85);
    \draw[gnuplot@green,pattern color=gnuplot@green!30,rounded corners=2pt,pattern=north east lines,] (3.15,3.15) rectangle (4.85,6.15);
    \draw[gnuplot@green,pattern color=gnuplot@green!30,rounded corners=2pt,pattern=north east lines,] (5.15,-0.15) rectangle (7.85,1.85);
    \draw[gnuplot@green,pattern color=gnuplot@green!30,rounded corners=2pt,pattern=north east lines,] (5.15,2.15) rectangle (7.85,3.85);
    \draw[gnuplot@green,pattern color=gnuplot@green!30,rounded corners=2pt,pattern=north east lines,] (5.15,4.15) rectangle (7.85,6.15);
    \draw[gnuplot@green,pattern color=gnuplot@green!30,rounded corners=2pt,pattern=north east lines,] (8.15,-0.15) rectangle (10.15,2.85);
    \draw[gnuplot@green,pattern color=gnuplot@green!30,rounded corners=2pt,pattern=north east lines,] (8.15,3.15) rectangle (10.15,6.15);
    \foreach \y in {0,1,2,3,4,5,6}
      \draw (0,\y) -- (10,\y);
    \foreach \x in {0,1,2,3,4,5,6,7,8,9,10}
      \draw (\x,0) -- (\x,6);
    \foreach \xx in {0,5}
      \foreach \x in {0,0.3,0.7,1,1.3,1.7,2,2.3,2.7,3,3.3,3.7,4,4.3,4.7,5}
        \foreach \y in {0,0.3,0.7,1,1.3,1.7,2,2.3,2.7,3,3.3,3.7,4,4.3,4.7,5,5.3,5.7,6}
          \draw (\xx+\x,\y) circle (0.06);
    \draw (5.1,-1) -- (5.1,-0.2);
    \draw (5.1,6.2) -- (5.1,7);
    \node[anchor=east] at (4.95,-0.6) {\footnotesize\sffamily process 0};
    \node[anchor=east] at (4.95,6.6) {\footnotesize\sffamily process 0};
    \node[anchor=west] at (5.25,-0.6) {\footnotesize\sffamily process 1};
    \node[anchor=west] at (5.25,6.6) {\footnotesize\sffamily process 1};
    \end{tikzpicture}
    \caption{Illustration of the numbering of degrees of freedom for a 2D
      setup with polynomial degree $p=3$. Cells are grouped together into
      batches of 6 cells and the interior unknowns are numbered first (highlighted by
      green-shaded boxes). The
      second set of numbers are unknowns located on more than one cell batch
      (not marked). The third set consists of unknowns that need to be exchanged with
      remote MPI processes (orange shades).
    }
  \label{fig:numbering}
\end{figure}

Given the mesh traversal, unknowns are enumerated in the sequence of the following four steps,
see also the illustration in Figure~\ref{fig:numbering}:
\begin{itemize}
  \setlength{\itemsep}{0pt}
  \setlength{\parskip}{0pt}
  \setlength{\parsep}{0pt}
\item In the first step, all unknowns touched only by a single cell batch are
  enumerated following the ordering of the cells. Except for reaching the next
  multiple of 64, this group will have a minimal distance of one between the ``pre''
  and ``post'' phase.
\item Next, the enumeration is continued on the unknowns touched by several
  batches, but not in contact with remote MPI processes. Here, some ranges
  will have a high distance, whereas others can still be completed reasonably
  close after the start.
\item In the third step, the unknowns owned locally, but requested by remote
  MPI ranks are assigned. These unknowns will not profit from overlap between
  the ``pre'' and ``post'' steps, because the ``pre'' step needs to be done
  before the initial \texttt{MPI\_Isend} command, whereas the ``post'' step comes
  after the final \texttt{MPI\_Recv} command. A contiguous numbering reduces
  the ranges of this unfavorable part of the vector to a minimum, besides also
  facilitating the pack/unpack operations.
\item Unknowns that are subject to constraints (not shown in
  Figure~\ref{fig:numbering}), such as Dirichlet boundary conditions, will not
  receive contributions from the matrix-free integrals with matrix
  $\myvector A$ representing a homogeneous operator. If they are kept in the
  linear system, like in the implementation of deal.II, they are appended at
  the end of the locally owned unknowns and updated during the last cell
  batch.
\end{itemize}
Furthermore, the numbering is set up to ensure contiguous numbers of multiple
unknowns associated with each vertex, edge, face, and volume, in order to
reduce the memory for index storage from $3(p+1)^3$ numbers per cell to $3^3$
numbers (for consistently oriented meshes). This reduces the memory
requirements of metadata, increases data locality and effectiveness of
prefetching as well as allows for packed load operations.

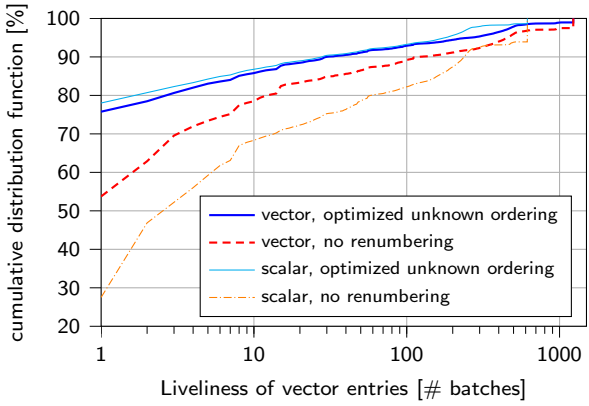
\begin{figure}
\centering
\begin{tikzpicture}
  \begin{semilogxaxis}[
    height=0.32\textwidth,
    width=0.45\textwidth,
    tick align=outside,
    tick pos=left,
    x grid style={white!69.0196078431373!black},
    xlabel={Liveliness of vector entries [\# batches]},
    xmajorgrids,
    xmin=1, xmax=1500,
    ymin=20, ymax=100,
    xtick style={color=black},
    y grid style={white!69.0196078431373!black},
    ylabel={cumulative distribution function [\%] },
    ymajorgrids,
    ytick distance=10,
    ytick style={color=black},
    legend cell align={left},
    xtick={1,10,100,1000},
    xticklabels={1,10,100,1000},legend pos = south east,
      legend style={/tikz/every even column/.append style={column sep=0.15cm},{font=\scriptsize\sffamily}}
    ]
    \addplot [thick, blue, mark options={solid,fill opacity=0}]
    table[x expr={\thisrowno{0}+1}] \tableLivelinessVectorRen;
    \addplot [thick, red, mark options={solid,fill opacity=0}, densely dashed]
    table[x expr={\thisrowno{0}+1}, y index=1] \tableLivelinessVectorDef;
    \addplot [cyan, mark options={solid,fill opacity=0}]
    table[x expr={\thisrowno{0}+1}] \tableLivelinessScalarRen;
    \addplot [orange, mark options={solid,fill opacity=0}, densely dashdotted]
    table[x expr={\thisrowno{0}+1}] \tableLivelinessScalarDef;
\legend{{vector, optimized unknown ordering},{vector, no renumbering},{scalar, optimized unknown ordering},{scalar, no renumbering}}

  \end{semilogxaxis}
\end{tikzpicture}
\caption{Liveliness of data in vector ranges for the 3D vector and scalar Laplacians with
  polynomial degree $p=5$ on 40 MPI processes. {The vector Laplacian involves 297 million DoFs subdivided into 1229 cell batches on each MPI process, the scalar Laplacian 99 million DoFs with 615 cell
  batches.}}
\label{fig:liveliness}
\end{figure}

Figure~\ref{fig:liveliness} visualizes the benefits of the proposed
{enumeration} algorithm by plotting the cumulative distribution function of the
liveliness of each subrange. We define ``liveliness'' as the number of cell
batches processed between the first and the last access, respectively. As a
reference, we also show the liveliness of the standard enumeration of degrees
of freedom in deal.II (enumeration in cell order).  The reduction of
the liveliness is clearly visible. {For the vector Laplacian, around 76\%---in contrast to 54\%}---of the
subranges is processed even in the same batch of cells. While the possibility
to process subranges within the same batch of cells is not necessary, we can
see similar trends for subranges living less than 10 batches of cells.  This
is an important threshold: For AVX-512 vectorization, each cell batch of 16
cells touches 12 kB of unique data (geometry, indices) for the matrix-vector
product and
\begin{align}\label{eq:data_access_between}
  &16\, \text{[cells]}\times  3 \cdot 5^3\, \text{[unique DoFs/cell]} \times 8\, \text{[byte/double]} \nonumber \\
  &\qquad = 48\, \text{[kB]}
\end{align}
of unique data per vector or 208 kB for four vectors and the preconditioner. For
around 10 cell batches, the data thus reaches the combined size of the L2 and
L3 caches per core on the Intel Skylake architecture. {For the scalar Laplacian, our heuristics use batches of 32 cells instead due to a lower number of DoFs/cell. This gives slightly better liveliness for our proposed numbering scheme, whereas the case with deal.II's default numbering scheme has even higher liveliness than in the vector case, as the DoFs with long liveliness are spread to many blocks of 64 DoFs when the number of unknowns per cell is lower.}

%
%
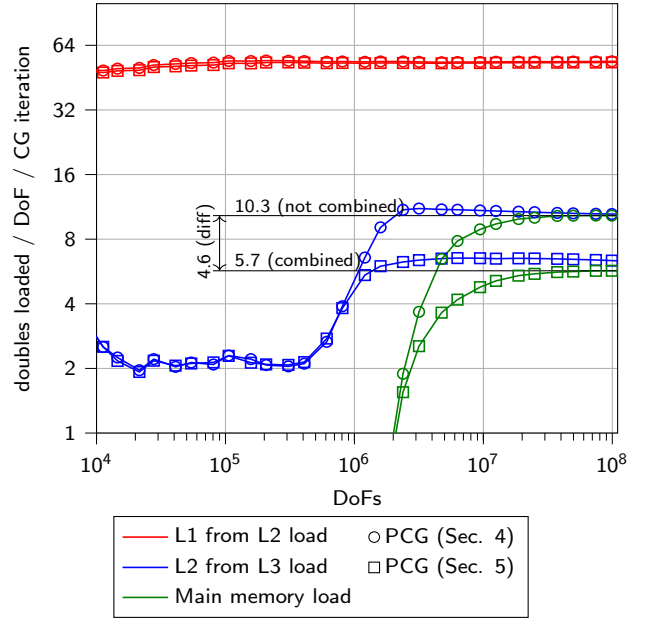
\begin{figure}
\centering
\begin{tikzpicture}
  \begin{loglogaxis}[
    log basis x={10},
    log basis y={2},
    tick align=outside,
    tick pos=left,
    x grid style={white!69.0196078431373!black},
    xlabel={DoFs},
    xmajorgrids,
    xmin=10000, xmax=110000000,
    xmode=log,
    xtick style={color=black},
    y grid style={white!69.0196078431373!black},
    ytick = {1,2,4,8,16,32,64},
    yticklabels={1,2,4,8,16,32,64},
    ylabel={{doubles loaded} / DoF / CG iteration },
    ymajorgrids,
    ymin=1, ymax=100,
    ytick style={color=black},
    legend cell align={left},
    legend to name=legend:likwidcaches,
    legend columns = 2,
    legend entries={L1 from L2 load,
      PCG (Sec. 4),
      L2 from L3 load,
      PCG (Sec. 5),
      Main memory load},
      legend style={/tikz/every even column/.append style={column sep=0.15cm}}
    ]
  \addlegendimage{no markers,red, semithick}
  \addlegendimage{only marks,mark=*, mark options={solid,fill opacity=0}}
  \addlegendimage{no markers,blue, semithick}
  \addlegendimage{only marks,mark=square}
  \addlegendimage{no markers,green!50.1960784313725!black, semithick}
  \addplot [semithick, blue, mark=*, mark size=2, mark options={solid,fill opacity=0}]
  table {%
  3168 16.3361896403055
  3993 7.37660586878101
  5808 10.547654246224
  7623 3.59981716515807
  11253 2.51809727616179
  14553 2.24837409711359
  21483 1.96218091554697
  27783 2.20908469207789
  41013 2.0344159169044
  54243 2.13045922976237
  80703 2.08596644486574
  105903 2.2972082943826
  157563 2.20923535347766
  206763 2.07393247341159
  307623 2.04491374182034
  408483 2.10504476318476
  610203 2.66079485023836
  807003 3.89632225654675
  1205523 6.56613353706234
  1594323 9.08057840224346
  2381643 10.9631381571461
  3168963 11.1109789700921
  4743603 11.0078905211924
  6298803 10.9720450060115
  9428643 10.882650345336
  12519843 10.8242970778467
  18740883 10.741337868125
  24961923 10.687631868346
  37404003 10.6010603664533
  49768803 10.5532268628201
  74575683 10.5030302739567
  99228483 10.4707300246644
  148688163 10.4358359800975
  };
  \addplot [semithick, blue, mark=square, mark size=2, mark options={solid,fill opacity=0}]
  table {%
  3168 16.5286631559497
  3993 7.49316623239605
  5808 10.6589751353662
  7623 3.56138495343041
  11253 2.52185002918141
  14553 2.16753368013872
  21483 1.92581495176166
  27783 2.17309145880575
  41013 2.06108429034696
  54243 2.10914311524068
  80703 2.13320756353543
  105903 2.28806077259379
  157563 2.12633200687979
  206763 2.08844183920721
  307623 2.07691313718415
  408483 2.14084796674525
  610203 2.75492336156984
  807003 3.81848797340282
  1205523 5.44416199442068
  1594323 5.98700435858982
  2381643 6.263900592994
  3168963 6.3897874478181
  4743603 6.49552407737325
  6298803 6.53157830146458
  9428643 6.52153761681294
  12519843 6.49097596511394
  18740883 6.51288182365793
  24961923 6.50262777230745
  37404003 6.47736904656969
  49768803 6.44518590290387
  74575683 6.40083896382149
  99228483 6.35740785737902
  148688163 6.30932117306473
  };
  \addplot [semithick, green!50.1960784313725!black, mark=*, mark size=2, mark options={solid,fill opacity=0}]
  table {%
  3168 4.16223977580685
  3993 1.99817766197228
  5808 2.36556889427187
  7623 0.922373081463991
  11253 0.615451495216891
  14553 0.533041499428054
  21483 0.491731075531917
  27783 0.458913724219847
  41013 0.539463097066784
  54243 0.295545047287208
  80703 0.447241738225345
  105903 0.290950209153659
  157563 0.386551728514943
  206763 0.298348834172458
  307623 0.366520058643209
  408483 0.196458604152437
  610203 0.205208348697073
  807003 0.158379212964512
  1205523 0.174379709055738
  1594323 0.416066098274942
  2381643 1.88887461302974
  3168963 3.67109445582041
  4743603 6.46675506993313
  6298803 7.84581610188475
  9428643 8.88688661242132
  12519843 9.41815174918727
  18740883 9.92753049576159
  24961923 10.0703416860151
  37404003 10.264919038211
  49768803 10.295237058444
  74575683 10.3077110772153
  99228483 10.303738846839
  148688163 10.2980213361033
  };
  \addplot [semithick, green!50.1960784313725!black, mark=square, mark size=2, mark options={solid,fill opacity=0}]
  table {%
  3168 4.57124599655087
  3993 1.83166285680792
  5808 2.42122933884298
  7623 0.915967712842713
  11253 0.660484531452273
  14553 0.56840918185456
  21483 0.509123492994461
  27783 0.465662455458374
  41013 0.547844585863019
  54243 0.35200394520952
  80703 0.449177849646234
  105903 0.343179607754266
  157563 0.369693392484276
  206763 0.301673897167288
  307623 0.35209493438397
  408483 0.203267333034667
  610203 0.218369952294564
  807003 0.141534480045303
  1205523 0.14197676029408
  1594323 0.390408656213327
  2381643 1.55015098820436
  3168963 2.53938551822789
  4743603 3.63776094879778
  6298803 4.1821686199743
  9428643 4.7806985586367
  12519843 5.11427389704487
  18740883 5.41006725777008
  24961923 5.51606265470813
  37404003 5.61985554727926
  49768803 5.65194676271398
  74575683 5.69784129566738
  99228483 5.70149172793461
  148688163 5.69165684022877
  };
  \addplot [semithick, red, mark=*, mark size=2, mark options={solid,fill opacity=0}]
  table {%
  3168 268.548672702636
  3993 51.5862866399177
  5808 285.463866723663
  7623 49.3917974386724
  11253 48.8045530200956
  14553 49.8078019086423
  21483 50.2166526009443
  27783 51.7638933880431
  41013 52.4246885133982
  54243 52.8864323507181
  80703 53.134641834876
  105903 54.1034602419195
  157563 54.027793644447
  206763 54.3142994636371
  307623 54.2008806233604
  408483 53.9903129383597
  610203 53.6480789999394
  807003 53.7778360179578
  1205523 53.5911591898288
  1594323 53.7960312935334
  2381643 53.7813864210547
  3168963 53.5833705694891
  4743603 53.3548549488648
  6298803 53.4828403110877
  9428643 53.5119058225028
  12519843 53.6218909654059
  18740883 53.6712304324188
  24961923 53.6809666566955
  37404003 53.6901182408204
  49768803 53.7403622104393
  74575683 53.7736166291632
  99228483 53.8692135780208
  148688163 53.9046484571203
  };
  \addplot [semithick, red, mark=square, mark size=2, mark options={solid,fill opacity=0}]
  table {%
  3168 270.329052722345
  3993 50.5205918868658
  5808 286.716226726513
  7623 48.4758297258297
  11253 47.7913097047995
  14553 48.6381421255371
  21483 49.0055078940072
  27783 50.6593510420041
  41013 51.0554580255041
  54243 51.4766882362701
  80703 51.7801382848221
  105903 52.7741187690623
  157563 52.7150251010707
  206763 53.1943577912876
  307623 53.2471970561369
  408483 53.1029596090902
  610203 52.8128036899196
  807003 52.9403623654435
  1205523 52.7558630154713
  1594323 52.9490024919668
  2381643 52.9149677344589
  3168963 52.8702331014909
  4743603 52.7707261758625
  6298803 52.877626114041
  9428643 52.896729147556
  12519843 53.0058888318328
  18740883 53.109912497186
  24961923 53.0972914226199
  37404003 53.1201165821744
  49768803 53.1939657057856
  74575683 53.172454686067
  99228483 53.2865815982494
  148688163 53.2738832999773
  };

	\addplot[mark=none, black] coordinates {(8e4,10.3) (2e8,10.3)};
	\addplot[mark=none, black] coordinates {(8e4,5.7) (2e8,5.7)};
	\draw[<->](axis cs:9e4,5.7)--(axis cs:9e4,10.3);

	\node[anchor=west] at (axis cs:1e5,11.3){\scriptsize\sffamily 10.3 (not combined)};
	\node[anchor=west] at (axis cs:1e5,6.3){\scriptsize\sffamily 5.7 (combined)};

	\node[anchor=east,rotate=90] at (axis cs:7e4,12.9){\scriptsize\sffamily 4.6 (diff)};

  \end{loglogaxis}
\end{tikzpicture}
\\
\strut\hfill\pgfplotslegendfromname{legend:likwidcaches}\hfill\strut
\caption{Comparison of measured memory transfer for $2\times 20$ cores of Intel Xeon Gold 6230
  for standard and combined versions of Algorithm 5.}
\label{fig:likwid_compare_mem_levels}
\end{figure}

While the consideration of liveliness is a rather theoretical approach of
quantifying the benefits for combining the ``pre'' operation, the
matrix-vector product, and the ``post'' operation, a clear reduction in the
data volume accessed from RAM can be observed.  A cache analysis (see
Figure~\ref{fig:likwid_compare_mem_levels}) conducted on the basis of hardware
performance counters using the LIKWID tool \citep{likwid10}
reveals that the combined version of {the PCG algorithm, as proposed here, only} reads 5.7
doubles per degree of freedom once the capacity of the caches is
exceeded. This value is {lower} by 4.6 doubles than the value of 10.3 reads for
the naive execution of Algorithm~\ref{alg:merged_precond_cg}. {
Note that the renumbering proposed above has further benefits beyond the
liveliness shown in Figure~\ref{fig:liveliness}, as the proposed scheme leads to a more linear data access pattern and fewer active streams, improving the effectiveness of hardware prefetching and reducing stress on the translation-lookaside buffers.}

\subsection{Comparison of CG variants}\label{sec:predict_access}

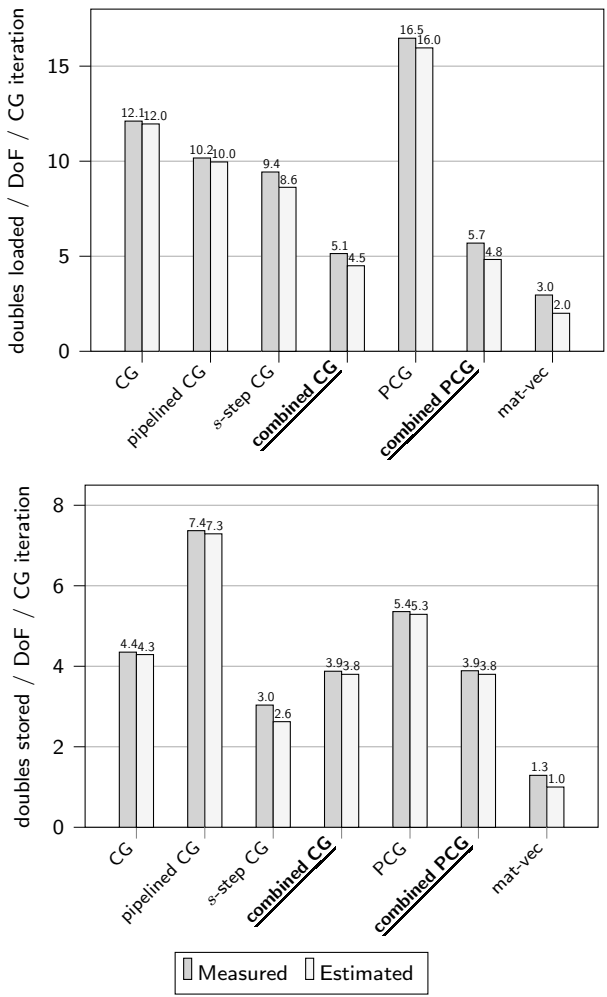
\begin{figure}
  \centering
  \begin{tikzpicture}
    \begin{axis}[
      ylabel={{doubles loaded} / DoF / CG iteration },
      height=0.72\columnwidth,
      width=0.98\columnwidth,
      tick align=outside,
      tick pos=left,
      x grid style={white!69.0196078431373!black},
      xmin=0, xmax=15,
      xtick style={color=black},
      xtick={1.5,3.5,5.5,7.5,9.5,11.5,13.5},
      xticklabels={CG, pipelined CG,{$s$-step CG},\underline{\smash{\textbf{combined CG}}},PCG,\underline{\smash{\textbf{combined PCG}}},mat-vec},
      xticklabel style={align=center,font=\scriptsize\sffamily, rotate=45,anchor=east},
      y grid style={white!69.0196078431373!black},
      ymin=0, ymax=18,
      ytick style={color=black},
      legend columns = 3,
      ymajorgrids,
      legend to name=legend:likwid,
      legend style={/tikz/every even column/.append style={column sep=0.15cm}}
      ]
      \draw[draw=black,fill=white!82.7450980392157!black] (axis cs:1,0) rectangle (axis cs:1.5,12.1094832041203);
      \addlegendimage{ybar,ybar legend,draw=black,fill=white!82.7450980392157!black};
      \addlegendentry{Measured}

      \draw[draw=black,fill=white!82.7450980392157!black] (axis cs:3,0) rectangle (axis cs:3.5,10.1682297752915);
      \draw[draw=black,fill=white!82.7450980392157!black] (axis cs:5,0) rectangle (axis cs:5.5,9.43147310061333);
      \draw[draw=black,fill=white!82.7450980392157!black] (axis cs:7,0) rectangle (axis cs:7.5,5.14151700663623);
      \draw[draw=black,fill=white!82.7450980392157!black] (axis cs:9,0) rectangle (axis cs:9.5,16.4698572037641);
      \draw[draw=black,fill=white!82.7450980392157!black] (axis cs:11,0) rectangle (axis cs:11.5,5.69165684022877);
      \draw[draw=black,fill=white!82.7450980392157!black] (axis cs:13,0) rectangle (axis cs:13.5,2.96011290421283);
      \draw[draw=black,fill=white!96.078431372549!black] (axis cs:1.5,0) rectangle (axis cs:2,11.9601129042128);
      \addlegendimage{ybar,ybar legend,draw=black,fill=white!96.078431372549!black};
      \addlegendentry{Estimated}

      \draw[draw=black,fill=white!96.078431372549!black] (axis cs:3.5,0) rectangle (axis cs:4,9.96011290421283);
      \draw[draw=black,fill=white!96.078431372549!black] (axis cs:5.5,0) rectangle (axis cs:6,8.62677957087949);
      \draw[draw=black,fill=white!96.078431372549!black] (axis cs:7.5,0) rectangle (axis cs:8,4.5);
      \draw[draw=black,fill=white!96.078431372549!black] (axis cs:9.5,0) rectangle (axis cs:10,15.9601129042128);
      \draw[draw=black,fill=white!96.078431372549!black] (axis cs:11.5,0) rectangle (axis cs:12,4.83);
      \draw[draw=black,fill=white!96.078431372549!black] (axis cs:13.5,0) rectangle (axis cs:14,2);
      \draw (axis cs:1.4,12.1601129042128) node[
        scale=0.5,
        anchor=base west,
        text=black,
        rotate=0.0
      ]{\sffamily 12.0};
      \draw (axis cs:3.4,10.1601129042128) node[
        scale=0.5,
        anchor=base west,
        text=black,
        rotate=0.0
      ]{\sffamily 10.0};
      \draw (axis cs:5.4,8.82677957087949) node[
        scale=0.5,
        anchor=base west,
        text=black,
        rotate=0.0
      ]{\sffamily 8.6};
      \draw (axis cs:7.4,4.7) node[
        scale=0.5,
        anchor=base west,
        text=black,
        rotate=0.0
      ]{\sffamily 4.5};
      \draw (axis cs:9.4,16.1601129042128) node[
        scale=0.5,
        anchor=base west,
        text=black,
        rotate=0.0
      ]{\sffamily 16.0};
      \draw (axis cs:11.4,5.03) node[
        scale=0.5,
        anchor=base west,
        text=black,
        rotate=0.0
      ]{\sffamily 4.8};
      \draw (axis cs:13.4,2.2) node[
        scale=0.5,
        anchor=base west,
        text=black,
        rotate=0.0
      ]{\sffamily 2.0};
      \draw (axis cs:0.75,12.3094832041203) node[
        scale=0.5,
        anchor=base west,
        text=black,
        rotate=0.0
      ]{\sffamily 12.1};
      \draw (axis cs:2.75,10.3682297752915) node[
        scale=0.5,
        anchor=base west,
        text=black,
        rotate=0.0
      ]{\sffamily 10.2};
      \draw (axis cs:4.9,9.63147310061333) node[
        scale=0.5,
        anchor=base west,
        text=black,
        rotate=0.0
      ]{\sffamily 9.4};
      \draw (axis cs:6.9,5.34151700663623) node[
        scale=0.5,
        anchor=base west,
        text=black,
        rotate=0.0
      ]{\sffamily 5.1};
      \draw (axis cs:8.9,16.6698572037641) node[
        scale=0.5,
        anchor=base west,
        text=black,
        rotate=0.0
      ]{\sffamily 16.5};
      \draw (axis cs:10.9,5.89165684022877) node[
        scale=0.5,
        anchor=base west,
        text=black,
        rotate=0.0
      ]{\sffamily 5.7};
      \draw (axis cs:12.9,3.16011290421283) node[
        scale=0.5,
        anchor=base west,
        text=black,
        rotate=0.0
      ]{\sffamily 3.0};
    \end{axis}
  \end{tikzpicture}
  \\
  \begin{tikzpicture}
    \begin{axis}[
      width=0.98\columnwidth,
      height=0.72\columnwidth,
      ylabel={{doubles stored} / DoF / CG iteration },
      tick align=outside,
      tick pos=left,
      x grid style={white!69.0196078431373!black},
      xmin=0, xmax=15,
      xtick={1.5,3.5,5.5,7.5,9.5,11.5,13.5},
      xticklabels={CG, pipelined CG,{$s$-step CG},\underline{\smash{\textbf{combined CG}}},PCG,\underline{\smash{\textbf{combined PCG}}},mat-vec},
      xticklabel style={align=center,font=\scriptsize\sffamily, rotate=45,anchor=east},
      y grid style={white!69.0196078431373!black},
      ymin=0, ymax=8.5,
      ytick style={color=black},
      ymajorgrids
      ]
      \draw[draw=black,fill=white!82.7450980392157!black] (axis cs:1,0) rectangle (axis cs:1.5,4.35101422633085);

      \draw[draw=black,fill=white!82.7450980392157!black] (axis cs:3,0) rectangle (axis cs:3.5,7.36887907983637);
      \draw[draw=black,fill=white!82.7450980392157!black] (axis cs:5,0) rectangle (axis cs:5.5,3.0359036324194);
      \draw[draw=black,fill=white!82.7450980392157!black] (axis cs:7,0) rectangle (axis cs:7.5,3.87679250734976);
      \draw[draw=black,fill=white!82.7450980392157!black] (axis cs:9,0) rectangle (axis cs:9.5,5.35811356079502);
      \draw[draw=black,fill=white!82.7450980392157!black] (axis cs:11,0) rectangle (axis cs:11.5,3.88893053645434);
      \draw[draw=black,fill=white!82.7450980392157!black] (axis cs:13,0) rectangle (axis cs:13.5,1.29035490202404);
      \draw[draw=black,fill=white!96.078431372549!black] (axis cs:1.5,0) rectangle (axis cs:2,4.29035490202404);

      \draw[draw=black,fill=white!96.078431372549!black] (axis cs:3.5,0) rectangle (axis cs:4,7.29035490202404);
      \draw[draw=black,fill=white!96.078431372549!black] (axis cs:5.5,0) rectangle (axis cs:6,2.62368823535738);
      \draw[draw=black,fill=white!96.078431372549!black] (axis cs:7.5,0) rectangle (axis cs:8,3.8);
      \draw[draw=black,fill=white!96.078431372549!black] (axis cs:9.5,0) rectangle (axis cs:10,5.29035490202404);
      \draw[draw=black,fill=white!96.078431372549!black] (axis cs:11.5,0) rectangle (axis cs:12,3.8);
      \draw[draw=black,fill=white!96.078431372549!black] (axis cs:13.5,0) rectangle (axis cs:14,1);
      \draw (axis cs:1.4,4.39035490202404) node[
        scale=0.5,
        anchor=base west,
        text=black,
        rotate=0.0
      ]{\sffamily 4.3};
      \draw (axis cs:3.4,7.39035490202404) node[
        scale=0.5,
        anchor=base west,
        text=black,
        rotate=0.0
      ]{\sffamily 7.3};
      \draw (axis cs:5.4,2.72368823535738) node[
        scale=0.5,
        anchor=base west,
        text=black,
        rotate=0.0
      ]{\sffamily 2.6};
      \draw (axis cs:7.4,3.9) node[
        scale=0.5,
        anchor=base west,
        text=black,
        rotate=0.0
      ]{\sffamily 3.8};
      \draw (axis cs:9.4,5.39035490202404) node[
        scale=0.5,
        anchor=base west,
        text=black,
        rotate=0.0
      ]{\sffamily 5.3};
      \draw (axis cs:11.4,3.9) node[
        scale=0.5,
        anchor=base west,
        text=black,
        rotate=0.0
      ]{\sffamily 3.8};
      \draw (axis cs:13.4,1.1) node[
        scale=0.5,
        anchor=base west,
        text=black,
        rotate=0.0
      ]{\sffamily 1.0};
      \draw (axis cs:0.9,4.45101422633085) node[
        scale=0.5,
        anchor=base west,
        text=black,
        rotate=0.0
      ]{\sffamily 4.4};
      \draw (axis cs:2.9,7.46887907983637) node[
        scale=0.5,
        anchor=base west,
        text=black,
        rotate=0.0
      ]{\sffamily 7.4};
      \draw (axis cs:4.9,3.1359036324194) node[
        scale=0.5,
        anchor=base west,
        text=black,
        rotate=0.0
      ]{\sffamily 3.0};
      \draw (axis cs:6.9,3.97679250734976) node[
        scale=0.5,
        anchor=base west,
        text=black,
        rotate=0.0
      ]{\sffamily 3.9};
      \draw (axis cs:8.9,5.45811356079502) node[
        scale=0.5,
        anchor=base west,
        text=black,
        rotate=0.0
      ]{\sffamily 5.4};
      \draw (axis cs:10.9,3.98893053645434) node[
        scale=0.5,
        anchor=base west,
        text=black,
        rotate=0.0
      ]{\sffamily 3.9};
      \draw (axis cs:12.9,1.39035490202404) node[
        scale=0.5,
        anchor=base west,
        text=black,
        rotate=0.0
      ]{\sffamily 1.3};

\end{axis}

\end{tikzpicture}
\\
\strut\hfill\pgfplotslegendfromname{legend:likwid}\hfill\strut
\caption{Comparison of measured and estimated memory transfer for various
  methods on $2\times 20$ cores of Intel Xeon Gold 6230  and $10^8$ DoFs, using
  the measured values of the matrix-vector multiplication as a baseline transfer.}
\label{fig:likwid_compare_mem}
\end{figure}

\begin{table}

\centering

\caption{Summary of the {modeled} ideal memory transfer of vector access regions
(see also the annotations in Algorithms~\ref{alg:precond_cg}-\ref{alg:merged_precond_cg}) and
matrix-vector multiplication for different CG variants.}
\label{tab:predicted_transfer}

\sffamily\sansmath\small
\begin{tabular}{lccccc}
\toprule
&\multicolumn{2}{c}{vector access} & & \multicolumn{2}{c}{mat-vec} \\
\cline{2-3}\cline{5-6}
& read & write & & read & write \\
\midrule
CG  & 9 & 3 & &2 & 1 \\
pipelined CG & 7 & 6 & &2 & 1 \\
$s$-step CG & 5 + 4/$s$ & 1 + 2/$s$ && 2 & 1 \\
\underline{\smash{\textbf{combined CG}}} & -- & -- & & 3.5 & 3.5\\
\midrule
PCG & 13  & 4 && 2 & 1 \\
\underline{\smash{\textbf{combined PCG}}} & -- & -- && 3.8$\overline{3}$ & 3.5 \\
\bottomrule
\end{tabular}
\end{table}

Since the reduction of the data volume to be transferred from/to main memory
is the key strength of the CG algorithm proposed in this work, we conclude
this section by comparing the measured read and write data volumes of the
basic CG, pipelined CG and $s$-step CG algorithms
(Algorithms~\ref{alg:precond_cg}--\ref{alg:merged_s_step}, {with loop fusion applied where possible}) with the results of
the proposed combined Algorithms~\ref{alg:merged_cg} and~\ref{alg:merged_precond_cg}.

Table~\ref{tab:predicted_transfer} shows the predicted memory read and write transfer
volumes in
different regions of the CG versions. In the
proposed combined CG variants, we assume that the reuse of memory reads
allows vectors to be read only once across the iteration of the algorithm
and, as a consequence, we do not separate estimates of the vector access regions
and of the matrix-vector product.

Figure~\ref{fig:likwid_compare_mem} presents both the estimated
and measured averaged
values of a complete iteration, derived from experiments with 100
iterations.  The cost of a single matrix-vector product (``mat-vec'') is 3.0
double data reads and 1.3 double data writes, which is slightly higher than
the theoretical expectation (2.0/1.0) due to non-perfect caching and loading of
the geometry data.  Since the optimization of the
memory transfer of this portion of the algorithm is not the focus of the
current work, we use the measured values of the
matrix-vector multiplication as a baseline transfer also for the CG algorithms.

In Figure~\ref{fig:likwid_compare_mem}, one can see that the measured values
match well with the predicted ones.  Furthermore, it is clear that while
the amount of data to be written by the combined versions is comparable with
the $s$-step version, they read up to 5 doubles less data from RAM compared
both to the pipelined and the $s$-step CG schemes.  Given the considerations
in the previous subsection, this improvement is expected, since a large
fraction of the vector entries accessed during the ``pre'' operation remains in
caches until they are read again during the ``post'' operation.

Compared to the theoretical transfer of 4.8 doubles per degree of freedom, the
excess transfer in the combined preconditioned method can be explained to a
good extent by the liveliness in Figure~\ref{fig:liveliness} and the
data-in-flight suggested by Equation~\eqref{eq:data_access_between}: 13\% of
vector entries have a liveliness of 10 or more cell batches, which can be
expected to give 2 additional reads between the ``pre'' operation and the
matrix-vector product as well as a transfer of $3.\overline{3}$ doubles to the
``post'' operation for the respective part of the vector. This explains 0.7
out of the 0.9 excess reads of doubles per DoF.

Furthermore, it is worth noting that the cost of the non-preconditioned and of the
Jacobi-preconditioned variant of the proposed CG algorithm is very close,
underlining that the benefit is even clearer in the preconditioned case.

In the next section, we evaluate the influence of
the reduced access to RAM and the reduced number of global reductions
on the throughput of CG algorithms for different scenarios
of high-order matrix-free finite-element methods.

\section{Numerical results}\label{sec:results}

In Sections~\ref{sec:improved_cg}--\ref{sec:interleave}, {we have proposed techniques
that reduce the access to main memory during CG iterations}. Since reducing the {memory access} is only a means to the application goal of increasing the
``throughput'', the CG and PCG variants for different numbers of compute
nodes are evaluated in the following, including some variations of the benchmark and the
behavior for different hardware. {Similarly to Section~\ref{sec:predict_access} basic loop fusion is applied in Algorithms~\ref{alg:precond_cg}--\ref{alg:merged_s_step} during evaluation.}

\subsection{Node-level performance}\label{sec:nodelevel}

Figure~\ref{fig:throughput_node} shows the throughput on a single compute node
for the different CG variants (no preconditioner) as well as the
preconditioned (PCG) case with a diagonal preconditioner. For small sizes, the
throughput is largely similar between the methods, given that the
matrix-vector product is the dominating cost and fast caches can absorb the
various vector access patterns.  As anticipated by the memory transfer
analysis from Section~\ref{sec:predict_access}, the picture changes when going to
larger sizes, where the {memory-transfer-efficient} combined variants are significantly faster.  The
advantage is particularly impressive considering that the {proposed} CG and PCG
variants, running at 2.36 and 2.13 billion DoFs/s for the largest sizes, are
separated from \emph{any} other method by a larger gap than what is observed
between the best and worst of the remaining schemes, the $s$-step CG method
($s=6$) with 1.46 billion DoFs/s and the preconditioned CG scheme with 0.98
billion DoFs/s.

\begin{figure}
  \begin{tikzpicture}
    \begin{semilogxaxis}[
      width=\columnwidth,
      height=0.75\columnwidth,
      xlabel={DoFs},
      ylabel={[billion DoFs $\times$ CG its] / [sec]},
      legend columns = 3,
      legend to name=legend:node,
      legend cell align={left},
      cycle list name=colorGPL,
      grid,
      semithick,
      ymin=0,ymax=3.4,ytick={0,0.5,1,1.5,2,2.5,3,3.5},
      xmin=1e4,xmax=1.6e8
      ]
      \addplot table[x expr={\thisrowno{3}}, y expr={\thisrowno{3}/min(\thisrowno{4},\thisrowno{12})*1e-9}] {\tableBasic};
      \addlegendentry{CG};
      \addplot table[x expr={\thisrowno{3}}, y expr={\thisrowno{3}/min(\thisrowno{4},\thisrowno{10})*1e-9}] {\tablePipelinedMerged};
      \addlegendentry{pipelined CG};
      \addplot table[x expr={\thisrowno{3}}, y expr={\thisrowno{3}/min(\thisrowno{4},\thisrowno{13})*1e-9}] {\tableSstepSix};
      \addlegendentry{$s$-step CG ($s=6$)};
      \addplot table[x expr={\thisrowno{3}}, y expr={\thisrowno{3}/min(\thisrowno{4},\thisrowno{8})*1e-9}] {\tableMerged};
      \addlegendentry{combined CG};
      \addplot table[x expr={\thisrowno{3}}, y expr={\thisrowno{3}/min(\thisrowno{4},\thisrowno{12})*1e-9}] {\tablePrec};
      \addlegendentry{PCG};
      \addplot table[x expr={\thisrowno{3}}, y expr={\thisrowno{3}/min(\thisrowno{4},\thisrowno{8})*1e-9}] {\tablePrecMerged};
      \addlegendentry{combined PCG};
      \addplot table[x index={1}, y expr={\thisrowno{1}*1e-9/min(\thisrowno{7},\thisrowno{13})}] {\tableMVDetailIntel};
      \addlegendentry{mat-vec};
    \end{semilogxaxis}
  \end{tikzpicture}
  \\
  \strut\hfill\pgfplotslegendfromname{legend:node}\hfill\strut
  \caption{Throughput over the problem size on a single Intel Xeon Platinum
    8174 node for the different CG variants.}
  \label{fig:throughput_node}
\end{figure}
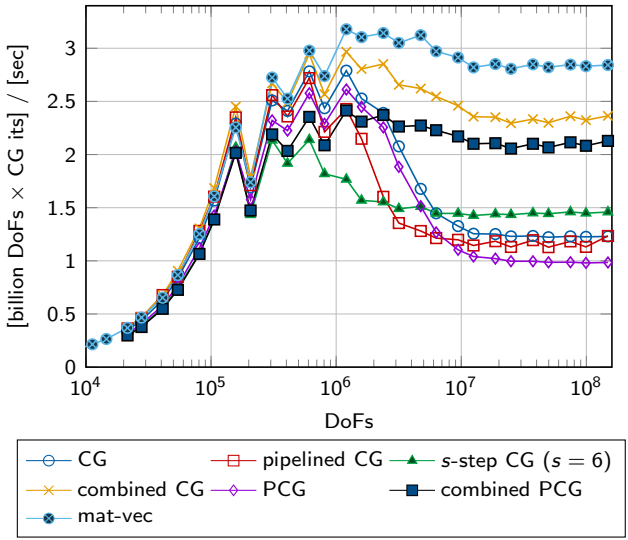

The mix of memory-intensive operations on vectors and the arithmetically heavy
matrix-vector product makes the throughput slightly deviate from the memory-transfer
predictions of Figure~\ref{fig:likwid_compare_mem}. For example, the
throughput of the combined CG scheme of 2.36 billion DoFs/s corresponds to an
average memory transfer of around 170 GB/s aggregated over the whole CG
solver, whereas the $s$-step method with 1.46 billion DoFs/s involves an
average transfer of 145 GB/s. While neither of the two variants saturates the
memory bandwidth on the present architecture, the achieved bandwidth
demonstrates an additional benefit of {our CG implementation} besides the
lower memory transfer: Fusing vector operations into an arithmetic-heavy
matrix-vector product allows to use spare memory bandwidth, leading to a
better distribution of the memory transfer.

Figure~\ref{fig:throughput_node} also shows the throughput of the
matrix-vector product alone as a point of reference. As its throughput is
20--30\% higher than that of the proposed merged variants, without the latter
fully saturating the available memory bandwidth, we suppose that further
performance improvements could be gained in the proposed algorithms by
suitable data prefetching. {Note that the slight oscillations in
  throughput are caused by differences in the amount of data exchange when
  cells are divisible by 48 or 64 as discussed before.}

\subsection{Scalability on up to 3,072 nodes}

Figure~\ref{fig:throughput_large} shows the throughput for different CG and
PCG variants on 512 compute nodes. The plot scales the achieved throughput by
the number of nodes, which allows a direct comparison with
Figure~\ref{fig:throughput_node}.  It can be seen that the behavior for large
sizes is similar to the single-node case, with a slight loss of around 5--10\%
in the parallel efficiency due to inter-node communication. For intermediate sizes
$n\sim 10^6$ per node, however, there is a pronounced difference.  In this regime, the
timings of a single iteration are in the range of the communication cost in
terms of a global reduction, canceling parts of the cache effect for those
intermediate sizes.

The experiments show that the proposed combined CG
variants achieve a similar performance for small sizes as the pipelined and
$s$-step methods, despite the latency optimization of the latter methods.
This can be explained by the number of \texttt{MPI\_Allreduce} calls per iteration,
which are one for both the combined CG method and the pipelined CG method
(albeit overlapped with the matrix-vector product for the latter), whereas the
$s$-step method with $s=6$ results in an average of 0.5 global reductions per
iteration.  The scaling limit becomes
even clearer when plotting the measured throughput over the time consumed by a
single CG iteration in the lower panel of Figure~\ref{fig:throughput_large},
directly showing the lowest possible iteration time. While the CG and PCG
methods are slower due to two and three global reductions per
iteration, respectively, all other methods take around $1.3\times 10^{-4}$ seconds as a
minimum time, which is caused by the global reduction combined with a scaling
limit of around $8\times 10^{-5}$ seconds for the matrix-vector product.

\begin{figure}
  \begin{tikzpicture}
    \begin{semilogxaxis}[
      width=\columnwidth,
      height=0.75\columnwidth,
      xlabel={DoFs / node},
      ylabel={[billion DoFs $\times$ CG its] / [nodes $\times$ sec]},
      legend columns = 3,
      legend to name=legend:large,
      legend cell align={left},
      cycle list name=colorGPL,
      grid,
      semithick,
      ymin=0,ymax=2.9,ytick={0,0.5,1,1.5,2,2.5,3},
      xmin=1e4,xmax=1.6e8
      ]
      \addplot table[x expr={\thisrowno{3}/512}, y expr={\thisrowno{3}/min(\thisrowno{4},\thisrowno{12})*1e-9/512}] {\tableLargeBasic};
      \addlegendentry{CG};
      \addplot table[x expr={\thisrowno{3}/512}, y expr={\thisrowno{3}/min(\thisrowno{4},\thisrowno{10})*1e-9/512}] {\tableLargePipelinedMerged};
      \addlegendentry{pipelined CG};
      \addplot table[x expr={\thisrowno{3}/512}, y expr={\thisrowno{3}/min(\thisrowno{4},\thisrowno{13})*1e-9/512}] {\tableLargeSstepSix};
      \addlegendentry{$s$-step CG ($s=6$)};
      \addplot table[x expr={\thisrowno{3}/512}, y expr={\thisrowno{3}/min(\thisrowno{4},\thisrowno{8})*1e-9/512}] {\tableLargeMerged};
      \addlegendentry{combined CG};
      \addplot table[x expr={\thisrowno{3}/512}, y expr={\thisrowno{3}/min(\thisrowno{4},\thisrowno{12})*1e-9/512}] {\tableLargePrec};
      \addlegendentry{PCG};
      \addplot table[x expr={\thisrowno{3}/512}, y expr={\thisrowno{3}/min(\thisrowno{4},\thisrowno{8})*1e-9/512}] {\tableLargePrecMerged};
      \addlegendentry{combined PCG};
      \addplot table[x expr={\thisrowno{3}/512}, y expr={\thisrowno{3}/min(\thisrowno{7},\thisrowno{9})*1e-9/512}] {\tableLargePrecMerged};
      \addlegendentry{mat-vec};
    \end{semilogxaxis}
  \end{tikzpicture}
  \\
  \begin{tikzpicture}
    \begin{semilogxaxis}[
      width=\columnwidth,
      height=0.75\columnwidth,
      xlabel={sec / CG it},
      ylabel={[billion DoFs $\times$ CG its] / [nodes $\times$ sec]},
      cycle list name=colorGPL,
      grid,
      semithick,
      ymin=0,ymax=2.9,ytick={0,0.5,1,1.5,2,2.5,3},
      xmin=5e-5,xmax=0.18
      ]
      \addplot table[x expr={min(\thisrowno{4},\thisrowno{12})}, y expr={\thisrowno{3}/min(\thisrowno{4},\thisrowno{12})*1e-9/512}] {\tableLargeBasic};
      \addplot table[x expr={min(\thisrowno{4},\thisrowno{10})}, y expr={\thisrowno{3}/min(\thisrowno{4},\thisrowno{10})*1e-9/512}] {\tableLargePipelinedMerged};
      \addplot table[x expr={min(\thisrowno{4},\thisrowno{13})}, y expr={\thisrowno{3}/min(\thisrowno{4},\thisrowno{13})*1e-9/512}] {\tableLargeSstepSix};
      \addplot table[x expr={min(\thisrowno{4},\thisrowno{8})}, y expr={\thisrowno{3}/min(\thisrowno{4},\thisrowno{8})*1e-9/512}] {\tableLargeMerged};
      \addplot table[x expr={min(\thisrowno{4},\thisrowno{12})}, y expr={\thisrowno{3}/min(\thisrowno{4},\thisrowno{12})*1e-9/512}] {\tableLargePrec};
      \addplot table[x expr={min(\thisrowno{4},\thisrowno{8})}, y expr={\thisrowno{3}/min(\thisrowno{4},\thisrowno{8})*1e-9/512}] {\tableLargePrecMerged};
      \addplot table[x expr={min(\thisrowno{7},\thisrowno{9})}, y expr={\thisrowno{3}/min(\thisrowno{7},\thisrowno{9})*1e-9/512}] {\tableLargePrecMerged};
    \end{semilogxaxis}
  \end{tikzpicture}
  \\
  \strut\hfill\pgfplotslegendfromname{legend:large}\hfill\strut
  \caption{Throughput over the problem size per node (top) and throughput over latency (bottom) on 512 nodes of Intel Xeon Platinum 8174 (24,576 MPI ranks) for various formulations.}
  \label{fig:throughput_large}
\end{figure}
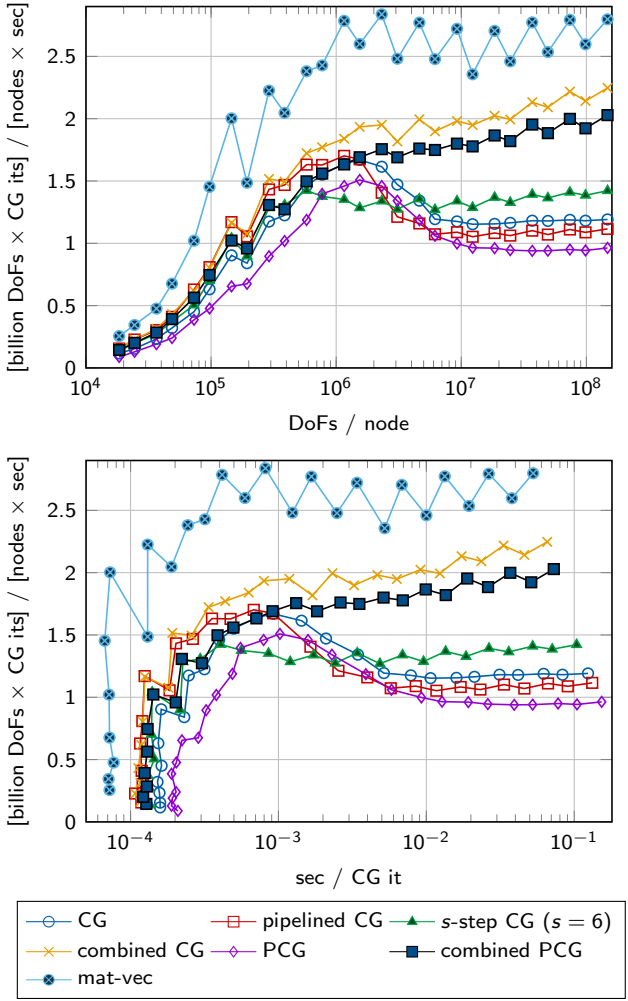

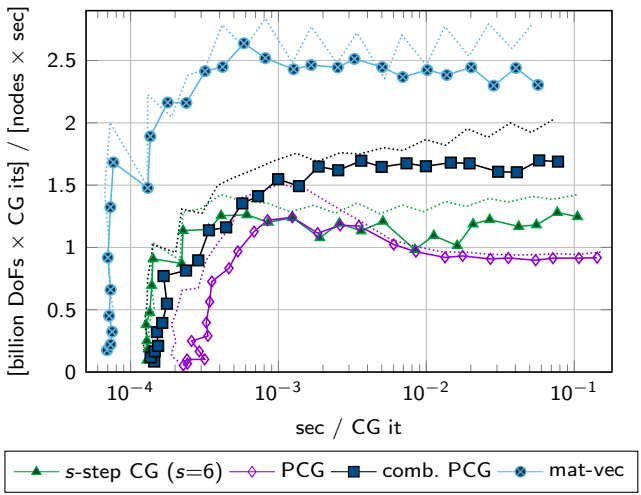
\begin{figure}
  \begin{tikzpicture}
    \begin{semilogxaxis}[
      width=\columnwidth,
      height=0.75\columnwidth,
      xlabel={sec / CG it},
      ylabel={[billion DoFs $\times$ CG its] / [nodes $\times$ sec]},
      cycle list name=colorGPL,
      legend columns = 4,
      legend to name=legend:largest,
      legend cell align={left},
      grid,
      semithick,
      ymin=0,ymax=2.9,ytick={0,0.5,1,1.5,2,2.5,3},
      xmin=5e-5,xmax=0.18
      ]
      \addplot[gnuplot@green,every mark/.append style={fill=gnuplot@green!50!black},mark=triangle*] table[x expr={\thisrowno{4}}, y expr={\thisrowno{3}/\thisrowno{4}*1e-9/3072}] {\tableLargestSstepSix};
      \addlegendentry{$s$-step CG ($s$=6)};
      \addplot[gnuplot@purple,mark=diamond] table[x expr={\thisrowno{4}}, y expr={\thisrowno{3}/\thisrowno{4}*1e-9/3072}] {\tableLargestPrec};
      \addlegendentry{PCG};
      \addplot[black,every mark/.append style={solid,fill=gnuplot@darkblue!80!black},mark=square*] table[x expr={min(\thisrowno{4},\thisrowno{8})}, y expr={\thisrowno{3}/min(\thisrowno{4},\thisrowno{8})*1e-9/3072}] {\tableLargestPrecMerged};
      \addlegendentry{comb.~PCG};
      \addplot[gnuplot@lightblue,every mark/.append style={solid,fill=gnuplot@lightblue!30!black},mark=otimes*] table[x expr={min(\thisrowno{7},\thisrowno{9})}, y expr={\thisrowno{3}/min(\thisrowno{7},\thisrowno{9})*1e-9/3072}] {\tableLargestPrecMerged};
      \addlegendentry{mat-vec};
      \addplot[densely dotted,gnuplot@green] table[x expr={min(\thisrowno{4},\thisrowno{13})}, y expr={\thisrowno{3}/min(\thisrowno{4},\thisrowno{13})*1e-9/512}] {\tableLargeSstepSix};
      \addplot[densely dotted,gnuplot@purple] table[x expr={min(\thisrowno{4},\thisrowno{12})}, y expr={\thisrowno{3}/min(\thisrowno{4},\thisrowno{12})*1e-9/512}] {\tableLargePrec};
      \addplot[densely dotted,black] table[x expr={min(\thisrowno{4},\thisrowno{8})}, y expr={\thisrowno{3}/min(\thisrowno{4},\thisrowno{8})*1e-9/512}] {\tableLargePrecMerged};
      \addplot[densely dotted,gnuplot@lightblue] table[x expr={min(\thisrowno{7},\thisrowno{9})}, y expr={\thisrowno{3}/min(\thisrowno{7},\thisrowno{9})*1e-9/512}] {\tableLargePrecMerged};
    \end{semilogxaxis}
  \end{tikzpicture}
  \\
  \strut\hfill\pgfplotslegendfromname{legend:largest}\hfill\strut
  \caption{Throughput over the problem size per node (top) and throughput over latency (bottom) on 3,072 nodes of Intel Xeon Platinum 8174 (147,456 MPI ranks) for various formulations. The dotted lines show the scaled throughput on 512 nodes (see also Figure~\ref{fig:throughput_large}).}
  \label{fig:throughput_largest}
\end{figure}

In Figure~\ref{fig:throughput_largest}, the throughput on 3,072 nodes against
the time of a single CG iteration is shown. By comparing with the result on
512 nodes (dotted lines), a slight loss in throughput for larger sizes can be
seen, corresponding to a small reduction in parallel efficiency for weak
scaling. Near the strong scaling limit in the left part of the figure, an
increase in the minimal time can be observed, which is due to the
higher cost of the global reductions on a larger scale. However, the increase
is similar between {the proposed combined} PCG algorithm and the baseline methods.
More importantly, the combined CG algorithm with preconditioner achieves a
throughput that is 35--40\% higher
than the one of the unpreconditioned $s$-step method for large sizes, confirming the
beneficial behavior of the proposed variant.

\subsection{Benchmark variations}

\newcommand{\plotbpsdegree}[3]{
  \begin{tikzpicture}
    \begin{semilogxaxis}[
      width=\columnwidth,
      height=0.75\columnwidth,
      xlabel={DoFs / node},
      ylabel={[billion DoFs $\times$ CG its] / [node $\times$ sec]},
      legend columns = 3,
      legend to name=legend:node,
      legend cell align={left},
      legend style={/tikz/every even column/.append style={column sep=0.2cm}},
      cycle list name=colorGPL,
      grid,
      semithick,
      ymin=0,ymax=2.4,ytick={0,0.5,1,1.5,2,2.5,3,3.5},
      xmin=1e4,xmax=5e8
      ]

      \addplot [red, dashed,mark=*] table[x expr={\thisrowno{3}/#1}, y expr={\thisrowno{3}/\thisrowno{4}*1e-9/#1}, col sep=comma] {ceed_results/bp4/node-#1-1.output};
      \addplot [red,mark=square] table[x expr={\thisrowno{3}/#1}, y expr={\thisrowno{3}/\thisrowno{7}*1e-9/#1}, col sep=comma] {ceed_results/bp4/node-#1-1.output};

      \addplot [gnuplot@darkblue, dashed,mark=*] table[x expr={\thisrowno{3}/#1}, y expr={\thisrowno{3}/\thisrowno{4}*1e-9/#1}, col sep=comma] {ceed_results/bp4/node-#1-3.output};
      \addplot [gnuplot@darkblue,mark=square]table[x expr={\thisrowno{3}/#1}, y expr={\thisrowno{3}/\thisrowno{7}*1e-9/#1}, col sep=comma] {ceed_results/bp4/node-#1-3.output};

      \addplot [gnuplot@orange, dashed,mark=*] table[x expr={\thisrowno{3}/#1}, y expr={\thisrowno{3}/\thisrowno{4}*1e-9/#1}, col sep=comma] {ceed_results/bp4/node-#1-5.output};
      \addplot [gnuplot@orange,mark=square]table[x expr={\thisrowno{3}/#1}, y expr={\thisrowno{3}/\thisrowno{7}*1e-9/#1}, col sep=comma] {ceed_results/bp4/node-#1-5.output};

      \addplot [gnuplot@purple, dashed,mark=*] table[x expr={\thisrowno{3}/#1}, y expr={\thisrowno{3}/\thisrowno{4}*1e-9/#1}, col sep=comma] {ceed_results/bp4/node-#1-7.output};
      \addplot [gnuplot@purple,mark=square]table[x expr={\thisrowno{3}/#1}, y expr={\thisrowno{3}/\thisrowno{7}*1e-9/#1}, col sep=comma] {ceed_results/bp4/node-#1-7.output};

      \addplot [gnuplot@green!50!black, dashed,mark=*] table[x expr={\thisrowno{3}/#1}, y expr={\thisrowno{3}/\thisrowno{4}*1e-9/#1}, col sep=comma] {ceed_results/bp4/node-#1-9.output};
      \addplot [gnuplot@green!50!black,mark=square]table[x expr={\thisrowno{3}/#1}, y expr={\thisrowno{3}/\thisrowno{7}*1e-9/#1}, col sep=comma] {ceed_results/bp4/node-#1-9.output};

    \end{semilogxaxis}
  \end{tikzpicture}
}

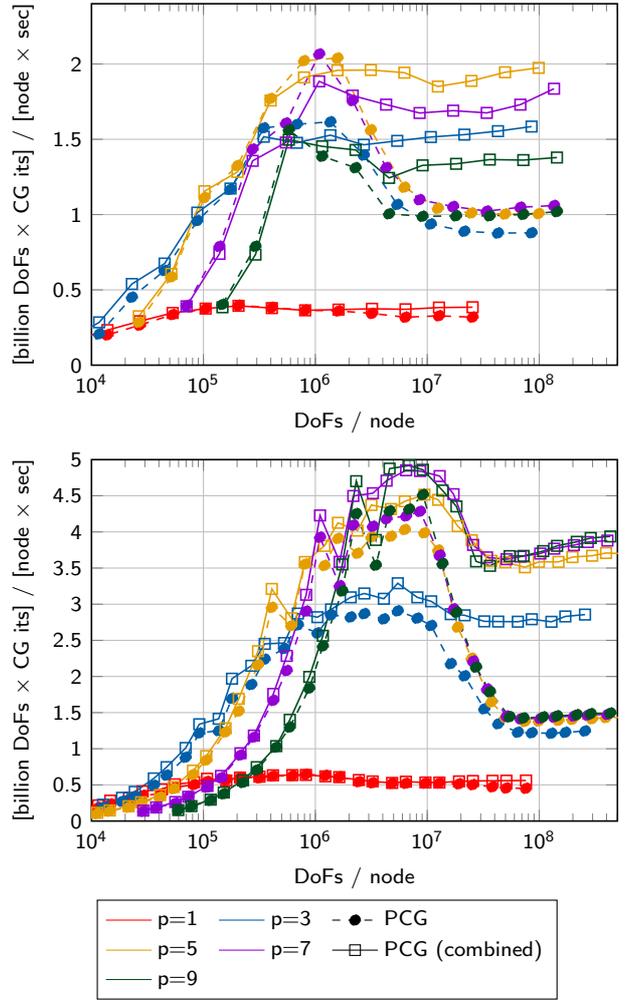
\begin{figure}

\plotbpsdegree{0004}{5}{4}
  \begin{tikzpicture}
    \begin{semilogxaxis}[
      width=\columnwidth,
      height=0.75\columnwidth,
      xlabel={DoFs / node},
      ylabel={[billion DoFs $\times$ CG its] / [node $\times$ sec]},
      legend columns = 3,
      legend to name=legend:node,
      legend cell align={left},
      legend style={/tikz/every even column/.append style={column sep=0.2cm}},
      cycle list name=colorGPL,
      grid,
      semithick,
      ymin=0,ymax=5,ytick={0,0.5,1,1.5,2,2.5,3,3.5, 4.0, 4.5, 5.0},
      xmin=1e4,xmax=5e8
      ]

  \addlegendimage{no markers,red}
  \addlegendentry{p=1};

  \addlegendimage{no markers,gnuplot@darkblue}
  \addlegendentry{p=3};

  \addlegendimage{dashed,mark=*}
  \addlegendentry{PCG};

  \addlegendimage{no markers,gnuplot@orange}
  \addlegendentry{p=5};

  \addlegendimage{no markers,gnuplot@purple}
  \addlegendentry{p=7};

  \addlegendimage{solid,mark=square}
  \addlegendentry{PCG (combined)};

  \addlegendimage{no markers,gnuplot@green!50!black}
  \addlegendentry{p=9};

      \addplot [red, dashed,mark=*] table[x expr={\thisrowno{3}}, y expr={\thisrowno{3}/\thisrowno{4}*1e-9}] {\tablePrecDegOne};
      \addplot [red,mark=square] table[x expr={\thisrowno{3}}, y expr={\thisrowno{3}/\thisrowno{4}*1e-9}] {\tablePrecMergedDegOne};

      \addplot [gnuplot@darkblue, dashed,mark=*] table[x expr={\thisrowno{3}}, y expr={\thisrowno{3}/\thisrowno{4}*1e-9}] {\tablePrecDegThree};
      \addplot [gnuplot@darkblue,mark=square]table[x expr={\thisrowno{3}}, y expr={\thisrowno{3}/\thisrowno{4}*1e-9}] {\tablePrecMergedDegThree};

      \addplot [gnuplot@orange, dashed,mark=*] table[x expr={\thisrowno{3}}, y expr={\thisrowno{3}/\thisrowno{4}*1e-9}] {\tablePrecDegFive};
      \addplot [gnuplot@orange,mark=square]table[x expr={\thisrowno{3}}, y expr={\thisrowno{3}/\thisrowno{4}*1e-9}] {\tablePrecMergedAMD};

      \addplot [gnuplot@purple, dashed,mark=*] table[x expr={\thisrowno{3}}, y expr={\thisrowno{3}/\thisrowno{4}*1e-9}] {\tablePrecDegSeven};
      \addplot [gnuplot@purple,mark=square]table[x expr={\thisrowno{3}}, y expr={\thisrowno{3}/\thisrowno{4}*1e-9}] {\tablePrecMergedDegSeven};

      \addplot [gnuplot@green!50!black, dashed,mark=*] table[x expr={\thisrowno{3}}, y expr={\thisrowno{3}/\thisrowno{4}*1e-9}] {\tablePrecDegNine};
      \addplot [gnuplot@green!50!black,mark=square]table[x expr={\thisrowno{3}}, y expr={\thisrowno{3}/\thisrowno{4}*1e-9}] {\tablePrecMergedDegNine};

    \end{semilogxaxis}
  \end{tikzpicture}
  \\
  \strut\hfill\pgfplotslegendfromname{legend:node}\hfill\strut
  \caption{Throughput over the problem size on
    four nodes of Intel Xeon Platinum 8174 (top) and
     on one node of $2\times 64$ core
    AMD Epyc 7742 (bottom) for the BP4 benchmark for different polynomial degrees $p$,
    all with quadratic geometry representation and $n_\text{q}=(p+2)^3$
    quadrature points.}
  \label{fig:throughput:poly}
\end{figure}

In the following, we will consider variants of the benchmark introduced in
Section~\ref{sec:matvec}. For the sake of simplicity and given the results
from the previous subsection, we concentrate on the basic PCG algorithm and
the proposed combined PCG algorithm.

\subsubsection{Variation of the polynomial degree}\label{sec:results:var:degree}

The results obtained for the polynomial degree $p=5$ above are transferable
also to other polynomials degrees, as shown in
Figure~\ref{fig:throughput:poly} (top panel). As examples, $p=1,3,5,7,9$ are
considered. {For} $p=1$, {the cost of computations
  compared to the number of unknowns is overwhelming}, as $(p+2)^3=27$
integration points {are used per cell compared to one unique
  unknown per cell, leading to a low throughput of 0.4 GDoFs/sec. This
  behavior specific to the present matrix-free operator evaluation behaves
  as $(p+2)^3/p^3$ and thus gives less work per unknown for higher degrees, as
  opposed to sparse matrix-vector products \citep{Kronbichler11,Kolev2021}. For
  higher degrees,} we can observe
significant speedups of the proposed PCG variants compared to the basic PCG,
with the highest throughput observed for $p=5$. Note that the maximal
achievable throughput decreases for the combined PCG algorithm as the
polynomial degree increases beyond $p\geq 7$, as opposed to constant
throughput for the basic PCG scheme.  This suggests that the fusion of vector
updates within matrix-vector product as proposed in
Section~\ref{sec:interleave} loses its benefits due to a limited cache
size. Caches need not only hold vector data of increasing size but also larger
temporary arrays for sum factorization \citep{Kronbichler2019}, with data of 8
elements in flight on the given AVX-512 hardware. Note that no tuning of the
parameters that have been identified in Section~\ref{sec:interleave} has been
performed, relying on simple heuristics.

\subsubsection{Variation of the geometric description}

According to the discussion in Section~\ref{sec:matvec:fast}, we have
concentrated on a tri-quadratic geometry description as a compromise between
higher-order geometry representation and high throughput up to now. However,
the beneficial behavior observed is transferable to other geometric descriptions as
well. Figure~\ref{fig:throughput_geometry} compares the proposed algorithm
with a basic PCG scheme on the two extrema of matrix-vector products from
Figure~\ref{fig:mv_compare}, one loading the inverse Jacobian at each
quadrature point and the other using an affine mesh with $n_q=p+1$ and
constant Jacobians.  While we observe a speedup of 2.17 in our base case of a
bilinear geometry description, the solver is 1.57$\times$ faster when loading
the inverse Jacobians and even 2.27$\times$ faster in the case of an affine
mesh. The relatively low improvement {when loading the inverse Jacobians} can be understood by
recalling the node-level performance analysis of Section~\ref{sec:nodelevel}
as the matrix-vector product is itself limited by the memory
bandwidth. Therefore, no additional memory transfer can be hidden behind
computations, reducing the advantage to the reduction in memory transfer only.

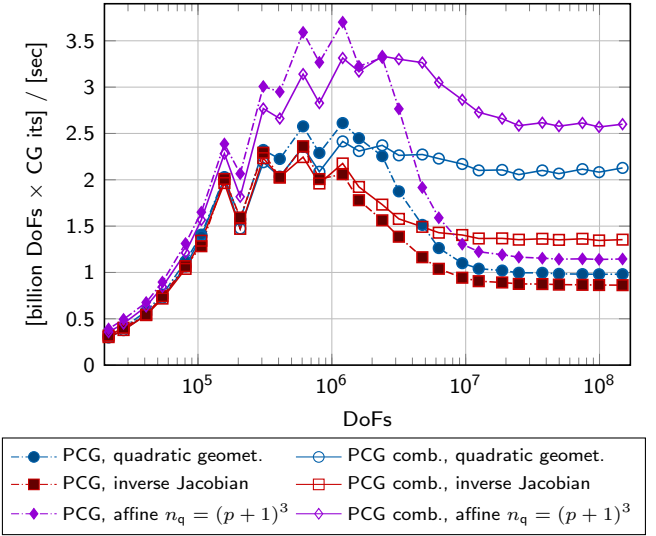
\begin{figure}
\centering
  \begin{tikzpicture}
    \begin{semilogxaxis}[
      width=\columnwidth,
      height=0.75\columnwidth,
      xlabel={DoFs},
      ylabel={[billion DoFs $\times$ CG its] / [sec]},
      legend cell align={left},
      cycle list name=colorGPL,
      legend to name=legendMVCompare,
      legend columns = 2,
      legend style={font=\scriptsize\sffamily},
      semithick,
      xmin=2e4,xmax=1.7e8,
      ymin=0,ymax=3.9,
      ytick={0,0.5,1,1.5,2,2.5,3,3.5},
      grid
      ]
      \addplot[gnuplot@darkblue,mark=*,densely dashdotted,every mark/.append style={solid,fill=gnuplot@darkblue!80!black}] table[x index={3}, y expr={\thisrowno{3}*1e-9/\thisrowno{4}}] {\tablePrec};
      \addlegendentry{PCG, quadratic geomet.};
      \addplot[gnuplot@darkblue,mark=o] table[x index={3}, y expr={\thisrowno{3}*1e-9/min(\thisrowno{4},\thisrowno{8})}] {\tablePrecMerged};
      \addlegendentry{PCG comb., quadratic geomet.};
      \addplot[gnuplot@red,every mark/.append style={fill=gnuplot@red!50!black,solid},mark=square*,densely dashdotted] table[x index={3}, y expr={\thisrowno{3}*1e-9/\thisrowno{4}}] {\tablePrecMVBasic};
      \addlegendentry{PCG, inverse Jacobian};
      \addplot[gnuplot@red,mark=square] table[x index={3}, y expr={\thisrowno{3}*1e-9/min(\thisrowno{4},\thisrowno{8})}] {\tablePrecMergedMVBasic};
      \addlegendentry{PCG comb., inverse Jacobian};
      \addplot[gnuplot@purple,mark=diamond*,densely dashdotted,every mark/.append style={solid}] table[x index={3}, y expr={\thisrowno{3}*1e-9/\thisrowno{4}}] {\tablePrecMVCart};
      \addlegendentry{PCG, affine $n_\text q=(p+1)^3$};
      \addplot[gnuplot@purple,mark=diamond] table[x index={3}, y expr={\thisrowno{3}*1e-9/min(\thisrowno{4},\thisrowno{8})}] {\tablePrecMergedMVCart};
      \addlegendentry{PCG comb., affine $n_\text q=(p+1)^3$};
    \end{semilogxaxis}
  \end{tikzpicture}
  \\
  \pgfplotslegendfromname{legendMVCompare}
  \caption{\sansmath Comparison of throughput of BP4 benchmark with basic
    preconditioned CG algorithm and the proposed combined variant with
    different implementations of matrix-free operator evaluation for
    polynomial degree $p=5$ on $2\times 24$ cores of Intel Xeon Platinum
    8174.}
  \label{fig:throughput_geometry}
\end{figure}

\subsubsection{Variation of the partial differential equation}

As a next set of tests,
we consider variants of the benchmark from \cite{Fischer19}, namely
BP1 (scalar mass matrix, $n_q=p+2$),
BP2 (vectorial mass matrix, $n_q=p+2$),
BP4 (scalar Laplace operator, $n_q=p+2$), and
BP5 (scalar Laplace operator, $n_q=p+1$, Gauss--Lobatto quadrature).
Figure~\ref{fig:throughput:bp} compares the throughput of the basic CG algorithm and of the
combined version for BP1--BP5. For large problem sizes ($\ge 5\times 10^6$ DoFs/node), a
clear trend is visible. While the throughput is limited to
0.8--1.2 GDoFs/sec for the basic CG scheme, the value is around two times
higher for the proposed algorithms with 1.8--2.3 GDoFs/sec for all cases. Also
note that the BP1, BP2, and BP5 cases with an arithmetically lighter
matrix-vector product saturate the RAM bandwidth with around 200 GB/s for the
combined CG iteration, whereas BP3 and BP4 reach around 170 GB/s bandwidth,
as seen above.

\newcommand{\plotbps}[3]{
\begin{figure}
  \begin{tikzpicture}
    \begin{semilogxaxis}[
      width=\columnwidth,
      height=0.75\columnwidth,
      xlabel={DoFs / node},
      ylabel={[billion DoFs $\times$ CG its] / [node $\times$ sec]},
      legend columns = 3,
      legend to name=legend:node,
      legend cell align={left},
      legend style={/tikz/every even column/.append style={column sep=0.2cm}},
      cycle list name=colorGPL,
      grid,
      semithick,
      ymin=0,ymax=3.9,ytick={0,0.5,1,1.5,2,2.5,3,3.5},
      xmin=1e4,xmax=1.6e8
      ]

  \addlegendimage{no markers,red}
  \addlegendentry{BP1};

  \addlegendimage{no markers,gnuplot@darkblue}
  \addlegendentry{BP2};

  \addlegendimage{dashed,mark=*}
  \addlegendentry{PCG};

  \addlegendimage{no markers,gnuplot@orange}
  \addlegendentry{BP3};

  \addlegendimage{no markers,gnuplot@purple}
  \addlegendentry{BP4};

  \addlegendimage{solid ,mark=square}
  \addlegendentry{PCG (combined)};

  \addlegendimage{no markers,gnuplot@green!50!black}
  \addlegendentry{BP5};

      \addplot [red, dashed,mark=*] table[x expr={\thisrowno{3}/#1}, y expr={\thisrowno{3}/\thisrowno{4}*1e-9/#1}, col sep=comma] {ceed_results/bp1/node-#1-#2.output};
      \addplot [red, mark=square] table[x expr={\thisrowno{3}/#1}, y expr={\thisrowno{3}/\thisrowno{8}*1e-9/#1}, col sep=comma] {ceed_results/bp1/node-#1-#2.output};

      \addplot [gnuplot@darkblue, dashed,mark=*] table[x expr={\thisrowno{3}/#1}, y expr={\thisrowno{3}/\thisrowno{4}*1e-9/#1}, col sep=comma] {ceed_results/bp2/node-#1-#2.output};
      \addplot [gnuplot@darkblue,mark=square]table[x expr={\thisrowno{3}/#1}, y expr={\thisrowno{3}/\thisrowno{7}*1e-9/#1}, col sep=comma] {ceed_results/bp2/node-#1-#2.output};

      \addplot [gnuplot@orange, dashed,mark=*] table[x expr={\thisrowno{3}/#1}, y expr={\thisrowno{3}/\thisrowno{4}*1e-9/#1}, col sep=comma] {ceed_results/bp3/node-#1-#2.output};
      \addplot [gnuplot@orange,mark=square]table[x expr={\thisrowno{3}/#1}, y expr={\thisrowno{3}/\thisrowno{7}*1e-9/#1}, col sep=comma] {ceed_results/bp3/node-#1-#2.output};

      \addplot [gnuplot@purple, dashed,mark=*] table[x expr={\thisrowno{3}/#1}, y expr={\thisrowno{3}/\thisrowno{4}*1e-9/#1}, col sep=comma] {ceed_results/bp4/node-#1-#2.output};
      \addplot [gnuplot@purple,mark=square]table[x expr={\thisrowno{3}/#1}, y expr={\thisrowno{3}/\thisrowno{7}*1e-9/#1}, col sep=comma] {ceed_results/bp4/node-#1-#2.output};

      \addplot [gnuplot@green!50!black, dashed,mark=*] table[x expr={\thisrowno{3}/#1}, y expr={\thisrowno{3}/\thisrowno{4}*1e-9/#1}, col sep=comma] {ceed_results/bp5/node-#1-#2.output};
      \addplot [gnuplot@green!50!black,mark=square]table[x expr={\thisrowno{3}/#1}, y expr={\thisrowno{3}/\thisrowno{7}*1e-9/#1}, col sep=comma] {ceed_results/bp5/node-#1-#2.output};

    \end{semilogxaxis}
  \end{tikzpicture}
  \\
  \strut\hfill\pgfplotslegendfromname{legend:node}\hfill\strut
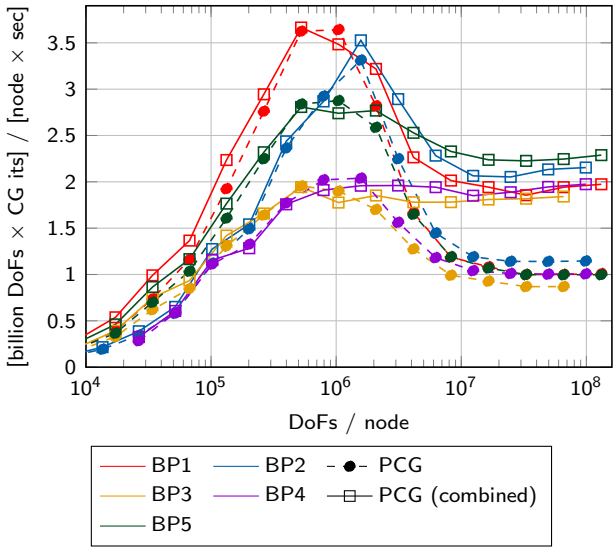
  \caption{Throughput over the problem size for the standard preconditioned CG
    scheme and the proposed improved version on #3~nodes of Intel Xeon
    Platinum 8174 for the CEED benchmark problems BP1 (scalar mass matrix),
    BP2 (vector-valued mass matrix), BP3 (scalar Laplace matrix), BP4
    (vector-valued Laplace matrix), and BP5 (scalar Laplace matrix,
    collocation setting with Gauss--Lobatto quadrature on $n_\text{q}=(p+1)^3$
    points) according to \cite{Fischer19}.}
  \label{fig:throughput:bp}
\end{figure}
}

\plotbps{0004}{5}{4}

\subsection{Comparison of different hardware}

In the following, we present results obtained on a dual-socket AMD Epyc 7742
CPU and a Nvidia Tesla V100 GPU. The AMD CPU consists of $2\times 64$ cores
running at 2.25 GHz and uses code compiled for the AVX2 instruction set
extension (4-wide SIMD). This gives an arithmetic peak performance of 4.61
TFlop/s. The memory configuration uses $2\times 8$ channels of DDR4-3200,
resulting {in} a peak bandwidth of 410 GB/s and a measured STREAM triad bandwidth
of 290 GB/s. The size of the last-level cache is 4 MB per core or 512 MB in
total. The Nvidia V100 provides an arithmetic peak performance of 7.8 TFlop/s,
a peak memory bandwidth of 900 GB/s, and a measured bandwidth of 720 GB/s. The
performance specifications of the V100 GPU are considerably higher on the GPU
compared to the two CPU systems, but with a less sophisticated cache
infrastructure.

\subsubsection{Variation of the polynomial degree on Intel and AMD CPUs}

The lower panel of Figure~\ref{fig:throughput:poly} shows the experiment from
Subsection~\ref{sec:results:var:degree}, varying the polynomial degree on an AMD
Epyc 7742 node. Here, we observe a maximal throughput of 4 GDoFs/sec and
a maximal speedups of 3$\times$ compared to the baseline CG solver
(compared to 2 GDoFs/sec and 2$\times$ speedup
in the case of Intel). This difference can be explained by the higher
arithmetic performance of the AMD system, shifting the performance limit with
an achieved bandwidth of around 270 GB/s closer to the memory throughput limit
of 290 GB/s. An interesting observation is the fact that the
performance does not drop for the high polynomial degrees $p>5$. This can be
contributed to larger caches as well as to the AVX-2 instruction-set
extension with vectorization aggregating work from only 4 cells together, which
increases the benefit of the combination of ``pre'', ``mat-vec'', and ``post''
regions.

\subsubsection{BP5 on CPU and GPU}

As a last experiment, we run {Algorithm~\ref{alg:merged_precond_cg}} on a GPU architecture.  Given
the much smaller available cache size compared to compute units, we
have not been able to embed the vector access regions into the cell-based
evaluation of the matrix-vector product. As a result, we propose to run the
three regions ``pre'', ``mat-vec'', and ``post'' each as a separate kernel
with its own kernel call. Furthermore, the matrix-vector product uses a
precomputed final coefficient on the GPU, due to a different balance between
arithmetic performance, available registers, and memory bandwidth compared to
CPUs, see also the analysis in \cite{Swirydowicz19}. Details on
the GPU infrastructure of deal.II can be found in \cite{Ljungkvist17} and
\cite{KronbichlerLjungkvist19}.

Figure~\ref{fig:gpu} shows the throughput of the regular and the combined
CG method run on a single GPU device on
Summit\footnote{\url{https://www.top500.org/system/179397/},
  retrieved on February 11, 2021.}(Nvidia V100). For small problem sizes,
a clear benefit can be observed due to the reduced number of kernel calls (3).
For large problem sizes, a speed-up of about 18\% with $2.8$ GDoF/s is
reached. Note that this represents a considerably lower improvement, which is
due to the missing overlap between the ``pre'' and ``post''
operations. Nonetheless, {Algorithm~\ref{alg:merged_precond_cg}} also improves the throughput for
lower sizes because of fewer kernel launches.
{Reducing the number of kernel calls in CG on GPUs has been also the motivation
in \cite{aliaga2013reformulated}, \cite{dehnavi2011enhancing},
\cite{rupp2016pipelined}, and \cite{Chalmers2020}. The contribution by \cite{rupp2016pipelined} was even able to obtain two kernel calls for
vector-matrix-multiplication implementations based on sparse matrices. However, the latter concept is not straightforwardly extensible to matrix-free finite-element computations with contributions to the result vector being accumulated from computations on several cells, as
discussed in Section~\ref{sec:dda}.}
Since the GPU's high-bandwidth memory is limited to 16 GB, the
maximum size of the problem that can be run is considerably smaller on the GPU.

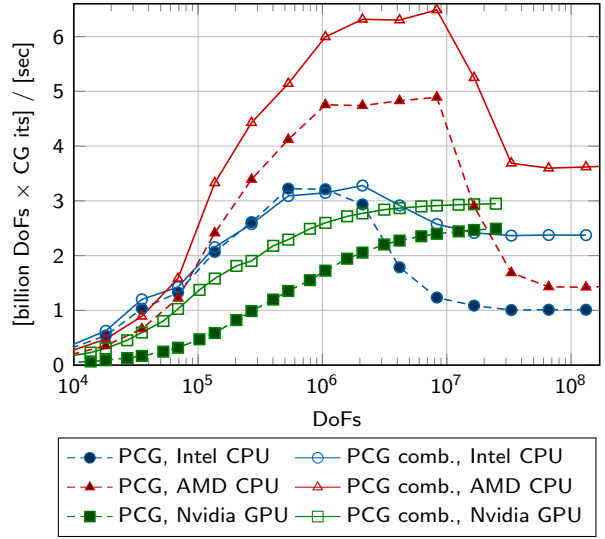
\begin{figure}
  \begin{tikzpicture}
    \begin{semilogxaxis}[
      width=\columnwidth,
      height=0.75\columnwidth,
      xlabel={DoFs},
      ylabel={[billion DoFs $\times$ CG its] / [sec]},
      legend columns = 2,
      legend to name=legend:node,
      legend cell align={left},
      cycle list name=colorGPL,
      grid,
      semithick,
      ymin=0,ymax=6.6,ytick={0,1,2,3,4,5,6},
      xmin=1e4,xmax=1.7e8
      ]

      \addplot [gnuplot@darkblue, densely dashed, mark=*,every mark/.append style={fill=gnuplot@darkblue!50!black,solid}] table[x expr={\thisrowno{3}}, y expr={\thisrowno{3}/\thisrowno{4}*1e-9}, col sep=comma] {ceed_results/bp5/node-0001-5.output};
      \addlegendentry{PCG, Intel CPU};
      \addplot [gnuplot@darkblue, solid, mark=o] table[x expr={\thisrowno{3}}, y expr={\thisrowno{3}/\thisrowno{7}*1e-9}, col sep=comma] {ceed_results/bp5/node-0001-5.output};
      \addlegendentry{PCG comb., Intel CPU};

      \addplot [gnuplot@red, densely dashed, mark=triangle*,every mark/.append style={fill=gnuplot@red!50!black,solid}] table[x expr={\thisrowno{3}}, y expr={\thisrowno{3}/\thisrowno{4}*1e-9}] {\tableBPFiveAMD};
      \addlegendentry{PCG, AMD CPU};
      \addplot [gnuplot@red, solid, mark=triangle] table[x expr={\thisrowno{3}}, y expr={\thisrowno{3}/\thisrowno{7}*1e-9}] {\tableBPFiveAMD};
      \addlegendentry{PCG comb., AMD CPU};

      \addplot [green!50!black, densely dashed, mark=square*,every mark/.append style={fill=green!30!black,solid}]
    table {%
1936	0.00911492
2541	0.0121402
3751	0.017973
4851	0.0232952
7161	0.0344083
9261	0.0445625
13671	0.0662658
18081	0.0887848
26901	0.129295
35301	0.167665
52521	0.246288
68921	0.317498
102541	0.471836
136161	0.586406
203401	0.822055
269001	0.986561
401841	1.19748
531441	1.35406
793881	1.55189
1056321	1.72487
1581201	1.94319
2099601	2.05532
3142881	2.20383
4173281	2.27445
6246961	2.3534
8320641	2.39855
12468001	2.44289
16589601	2.46848
24858561	2.49033
};
\addlegendentry{PCG, Nvidia GPU};

      \addplot [green!50!black, solid, mark=square] table{
1936	0.0302771
2541	0.044566
3751	0.0705031
4851	0.088201
7161	0.130642
9261	0.16489
13671	0.229745
18081	0.304692
26901	0.453979
35301	0.59578
52521	0.805754
68921	1.0265
102541	1.37388
136161	1.58157
203401	1.81239
269001	1.90373
401841	2.17913
531441	2.29201
793881	2.48971
1056321	2.5981
1581201	2.71785
2099601	2.77161
3142881	2.83913
4173281	2.86701
6246961	2.89821
8320641	2.91254
12468001	2.92952
16589601	2.9381
24858561	2.94795
};
\addlegendentry{PCG comb., Nvidia GPU};

    \end{semilogxaxis}
  \end{tikzpicture}
  \\
  \strut\hfill\pgfplotslegendfromname{legend:node}\hfill\strut
  \caption{BP5: Throughput over the problem size on a single node for the
    basic preconditioned CG method and the proposed combined variant.}
  \label{fig:gpu}
\end{figure}

With the proposed combined CG method, the CPU results appear more
beneficial than the GPU results: Given that both the memory bandwidth and
arithmetic performance is considerably higher on a single V100 device than on the
dual-socket Intel and AMD systems, one would expect best performance on the
GPU. However, the Intel result is only 20\% lower than the GPU, and the AMD
result from RAM is 20\% better than on the GPU, because of the reduction of
memory transfer between the ``pre'' and ``post'' regions. Furthermore, the CPU
reach a higher throughput for moderate sizes when the data fits into caches.

\section{Conclusions}\label{sec:conclusions}

We have presented {an implementation} of the conjugate gradient method that aims to
minimize the access to auxiliary vectors for the case of high-order
matrix-free finite-element implementations with a diagonal preconditioner. The
development was motivated by the observation that matrix-free operator
evaluation has become so fast that \texttt{AXPY}-style vector
updates, dot products and the application of the preconditioner can consume
around two thirds of the total runtime for large problem sizes on modern hardware, relevant for
example for fluid dynamics applications. The proposed solver relies on
interleaving the vector updates and dot products of the conjugate gradient
iteration with the loop through the mesh elements of the matrix-vector product,
combined with redundant applications of the preconditioner and summation of
auxiliary quantities to break the dependencies. We have
shown that around 90\% of the vector entries in the three active vectors of a
CG iteration can be re-used from fast cache memory,
resulting in a single load and store operation for each vector.

Both node-level performance analyses and strong/weak-scaling studies on up to
147,456 CPU cores confirm the suitability of the proposed algorithm for modern
hardware. Experiments have been conducted on CPU-based (Intel Xeon Platinum
8174, AMD Epyc 7742) and GPU-based (Nvidia Tesla V100 GPU) compute nodes for a
large variety of polynomial degrees, geometric descriptions, and PDEs
(scalar/vector-valued mass/Laplace matrix). Compared to a baseline CG solver
as well as optimized pipelined CG and $s$-step CG implementations, speedups of
2--3$\times$ have been reported. Besides reducing the memory transfer, the
proposed method allows to run memory-heavy vector operations near the
arithmetic-heavy matrix-free operator evaluation. As a result, new tuning
opportunities for implementing matrix-free methods appear, allowing to
gain performance from computing, e.g., redundant geometry information on the fly
with reduced memory transfer, an operation that might not be beneficial for
the matrix-vector product alone.

Future work aims to extend the algorithm towards the data dependencies imposed
by discontinuous Galerkin discretizations as well as more
sophisticated preconditioners {with longer-range data dependencies}. Furthermore, it would be useful to apply
analysis and transformation tools from compiler constructions to replace the
current manual dependency management for interleaving the matrix-vector
product with vector updates and inner products by a more automatic
approach based on hardware characteristics, which would make the application
to other algorithms, like BiCGStab or GMRES, simpler.

\footnotesize
\section*{Acknowledgements}
The authors acknowledge collaboration with
Momme Allalen,
Daniel Arndt,
Paddy \'O Conbhu\'i,
Prashanth Kanduri,
Karl Ljungkvist,
Alexander Roschlaub,
Bruno Turcksin,
as well as the deal.II community.

This work was supported by the Bayerisches Kompetenznetzwerk für
Technisch-Wissenschaftliches Hoch- und Höchstleistungsrechnen (KONWIHR)
through the projects ``Performance tuning of high-order discontinuous Galerkin
solvers for SuperMUC-NG'' and ``High-order matrix-free finite element
implementations with hybrid parallelization and improved data locality''. The
authors gratefully acknowledge the Gauss Centre for Supercomputing e.V.
(\url{www.gauss-centre.eu}) for funding this project by providing computing
time on the GCS Supercomputer SuperMUC-NG at Leibniz Supercomputing Centre
(LRZ, \url{www.lrz.de}) through project id pr83te.

\bibliographystyle{abbrvnat}
\bibliography{paper}

\begin{thebibliography}{49}
\providecommand{\natexlab}[1]{#1}
\providecommand{\url}[1]{\texttt{#1}}
\expandafter\ifx\csname urlstyle\endcsname\relax
  \providecommand{\doi}[1]{doi: #1}\else
  \providecommand{\doi}{doi: \begingroup \urlstyle{rm}\Url}\fi

\bibitem[Aliaga et~al.(2013)Aliaga, P{\'e}rez, Quintana-Ort{\'\i}, and
  Anzt]{aliaga2013reformulated}
J.~I. Aliaga, J.~P{\'e}rez, E.~S. Quintana-Ort{\'\i}, and H.~Anzt.
\newblock Reformulated conjugate gradient for the energy-aware solution of
  linear systems on {GPUs}.
\newblock In \emph{2013 42nd International Conference on Parallel Processing},
  pages 320--329. IEEE, 2013.

\bibitem[Arndt et~al.(2020{\natexlab{a}})Arndt, Bangerth, Blais, Clevenger,
  Fehling, Grayver, Heister, Heltai, Kronbichler, Maier, Munch, Pelteret,
  Rastak, Tomas, Turcksin, Wang, and Wells]{dealii92}
D.~Arndt, W.~Bangerth, B.~Blais, T.~C. Clevenger, M.~Fehling, A.~V. Grayver,
  T.~Heister, L.~Heltai, M.~Kronbichler, M.~Maier, P.~Munch, J.-P. Pelteret,
  R.~Rastak, I.~Tomas, B.~Turcksin, Z.~Wang, and D.~Wells.
\newblock The deal.{II} library, version 9.2.
\newblock \emph{Journal of Numerical Mathematics}, 28\penalty0 (3):\penalty0
  131--146, 2020{\natexlab{a}}.
\newblock \doi{10.1515/jnma-2020-0043}.
\newblock URL \url{https//dealii.org}.

\bibitem[Arndt et~al.(2020{\natexlab{b}})Arndt, Fehn, Kanschat, Kormann,
  Kronbichler, Munch, Wall, and Witte]{Arndt20}
D.~Arndt, N.~Fehn, G.~Kanschat, K.~Kormann, M.~Kronbichler, P.~Munch, W.~A.
  Wall, and J.~Witte.
\newblock {ExaDG} -- high-order discontinuous {G}alerkin for the exa-scale.
\newblock In H.-J. Bungartz, S.~Reiz, B.~Uekermann, P.~Neumann, and W.~E.
  Nagel, editors, \emph{Software for Exascale Computing -- SPPEXA 2016--2019},
  Lecture Notes in Computational Science and Engineering 136, pages 189--224,
  Cham, 2020{\natexlab{b}}. Springer International Publishing.
\newblock \doi{10.1007/978-3-030-47956-5_8}.

\bibitem[Arndt et~al.(2021)Arndt, Bangerth, Davydov, Heister, Heltai,
  Kronbichler, Maier, Pelteret, Turcksin, and Wells]{dealiicanonical}
D.~Arndt, W.~Bangerth, D.~Davydov, T.~Heister, L.~Heltai, M.~Kronbichler,
  M.~Maier, J.-P. Pelteret, B.~Turcksin, and D.~Wells.
\newblock The deal.{{II}} finite element library: design, features, and
  insights.
\newblock \emph{Computers \& Mathematics with Applications}, 81:\penalty0
  407--422, 2021.
\newblock \doi{10.1016/j.camwa.2020.02.022}.

\bibitem[Bangerth et~al.(2011)Bangerth, Burstedde, Heister, and
  Kronbichler]{Bangerth11}
W.~Bangerth, C.~Burstedde, T.~Heister, and M.~Kronbichler.
\newblock Algorithms and data structures for massively parallel generic
  adaptive finite element codes.
\newblock \emph{ACM Transactions on Mathematical Software}, 38:\penalty0
  14/1--28, 2011.
\newblock \doi{10.1145/2049673.2049678}.

\bibitem[Bauer et~al.(2018)Bauer, Drzisga, Mohr, R\"{u}de, Waluga, and
  Wohlmuth]{Bauer2018}
S.~Bauer, D.~Drzisga, M.~Mohr, U.~R\"{u}de, C.~Waluga, and B.~Wohlmuth.
\newblock A stencil scaling approach for accelerating matrix-free finite
  element implementations.
\newblock \emph{{SIAM} Journal on Scientific Computing}, 40\penalty0
  (6):\penalty0 C748--C778, 2018.
\newblock \doi{10.1137/17m1148384}.

\bibitem[Brown(2010)]{Brown10}
J.~Brown.
\newblock Efficient nonlinear solvers for nodal high-order finite elements in
  {3D}.
\newblock \emph{Journal of Scientific Computing}, 45\penalty0 (1-3):\penalty0
  48--63, 2010.
\newblock \doi{10.1007/s10915-010-9396-8}.

\bibitem[Burstedde et~al.(2011)Burstedde, Wilcox, and Ghattas]{Burstedde11}
C.~Burstedde, L.~C. Wilcox, and O.~Ghattas.
\newblock p4est: {S}calable algorithms for parallel adaptive mesh refinement on
  forests of octrees.
\newblock \emph{SIAM J. Sci. Comput.}, 33\penalty0 (3):\penalty0 1103--1133,
  2011.
\newblock \doi{10.1137/10079163}.
\newblock URL \url{http://p4est.org}.

\bibitem[Chalmers and Warburton(2020)]{Chalmers2020}
N.~Chalmers and T.~Warburton.
\newblock Portable high-order finite element kernels {I}: Streaming operations,
  2020.
\newblock arXiv:2009.10917 preprint.

\bibitem[Charrier et~al.(2019)Charrier, Hazelwood, Tutlyaeva, Bader, Dumbser,
  Kudryavtsev, Moskovsky, and Weinzierl]{Charrier19}
D.~E. Charrier, B.~Hazelwood, E.~Tutlyaeva, M.~Bader, M.~Dumbser,
  A.~Kudryavtsev, A.~Moskovsky, and T.~Weinzierl.
\newblock Studies on the energy and deep memory behaviour of a cache-oblivious,
  task-based hyperbolic {PDE} solver.
\newblock \emph{The International Journal of High Performance Computing
  Applications}, 33\penalty0 (5):\penalty0 973--986, 2019.
\newblock \doi{10.1177/1094342019842645}.

\bibitem[Chronopoulos and Gear(1989)]{sstep1989}
A.~T. Chronopoulos and C.~W. Gear.
\newblock S-step iterative methods for symmetric linear systems.
\newblock \emph{Journal of Computational and Applied Mathematics}, 25\penalty0
  (2):\penalty0 153--168, Feb. 1989.
\newblock ISSN 0377-0427.
\newblock \doi{10.1016/0377-0427(89)90045-9}.

\bibitem[Cornelis et~al.(2018)Cornelis, Cools, and Vanroose]{pipelined18}
J.~Cornelis, S.~Cools, and W.~Vanroose.
\newblock The communication-hiding conjugate gradient method with deep
  pipelines.
\newblock \emph{ArXiv e-prints}, 1801.4728v3, 2018.

\bibitem[Dehnavi et~al.(2011)Dehnavi, Fern{\'a}ndez, and
  Giannacopoulos]{dehnavi2011enhancing}
M.~M. Dehnavi, D.~M. Fern{\'a}ndez, and D.~Giannacopoulos.
\newblock Enhancing the performance of conjugate gradient solvers on graphic
  processing units.
\newblock \emph{IEEE Transactions on Magnetics}, 47\penalty0 (5):\penalty0
  1162--1165, 2011.

\bibitem[Demkowicz et~al.(1990)Demkowicz, Oden, and Rachowicz]{Demkowicz90}
L.~Demkowicz, J.~Oden, and W.~Rachowicz.
\newblock A new finite element method for solving compressible
  {N}avier--{S}tokes equations based on an operator splitting method and h-p
  adaptivity.
\newblock \emph{Computer Methods in Applied Mechanics and Engineering},
  84\penalty0 (3):\penalty0 275--326, 1990.
\newblock \doi{10.1016/0045-7825(90)90081-v}.

\bibitem[Deville et~al.(2002)Deville, Fischer, and Mund]{Deville02}
M.~O. Deville, P.~F. Fischer, and E.~H. Mund.
\newblock \emph{High-order methods for incompressible fluid flow}, volume~9.
\newblock Cambridge University Press, 2002.

\bibitem[Eisenstat(1981)]{Eisenstat81}
S.~C. Eisenstat.
\newblock Efficient implementation of a class of preconditioned conjugate
  gradient methods.
\newblock \emph{{SIAM} Journal on Scientific and Statistical Computing},
  2\penalty0 (1):\penalty0 1--4, 1981.
\newblock \doi{10.1137/0902001}.

\bibitem[Eller et~al.(2019)Eller, Hoefler, and Gropp]{Eller2019}
P.~R. Eller, T.~Hoefler, and W.~Gropp.
\newblock Using performance models to understand scalable {K}rylov solver
  performance at scale for structured grid problems.
\newblock In \emph{Proceedings of the {ACM} International Conference on
  Supercomputing}. {ACM}, June 2019.
\newblock \doi{10.1145/3330345.3330358}.
\newblock URL \url{https://doi.org/10.1145/3330345.3330358}.

\bibitem[Fehn et~al.(2018)Fehn, Wall, and Kronbichler]{Fehn18}
N.~Fehn, W.~A. Wall, and M.~Kronbichler.
\newblock Efficiency of high-performance discontinuous {G}alerkin spectral
  element methods for under-resolved turbulent incompressible flows.
\newblock \emph{International Journal for Numerical Methods in Fluids},
  88\penalty0 (1):\penalty0 32--54, 2018.
\newblock \doi{10.1002/fld.4511}.

\bibitem[Fischer et~al.(2020)Fischer, Min, Rathnayake, Dutta, Kolev, Dobrev,
  Camier, Kronbichler, Warburton, {\'{S}}wirydowicz, and Brown]{Fischer19}
P.~Fischer, M.~Min, T.~Rathnayake, S.~Dutta, T.~Kolev, V.~Dobrev, J.-S. Camier,
  M.~Kronbichler, T.~Warburton, K.~{\'{S}}wirydowicz, and J.~Brown.
\newblock Scalability of high-performance {PDE} solvers.
\newblock \emph{The International Journal of High Performance Computing
  Applications}, 34\penalty0 (5):\penalty0 562--586, 2020.
\newblock \doi{10.1177/1094342020915762}.

\bibitem[Fischer et~al.(2021)Fischer, Kerkemeier, Peplinski, Shaver,
  Tomboulides, Min, Obabko, and Merzari]{nek5000}
P.~Fischer, S.~Kerkemeier, A.~Peplinski, D.~Shaver, A.~Tomboulides, M.~Min,
  A.~Obabko, and E.~Merzari.
\newblock Nek5000 {W}eb page.
\newblock 2021.
\newblock {https://nek5000.mcs.anl.gov}.

\bibitem[Ghysels and Vanroose(2014)]{pipelined14}
P.~Ghysels and W.~Vanroose.
\newblock Hiding global synchronization latency in the preconditioned conjugate
  gradient algorithm.
\newblock \emph{Parallel Computing}, 40\penalty0 (7):\penalty0 224--238, 2014.
\newblock \doi{10.1016/j.parco.2013.06.001}.
\newblock 7th Workshop on Parallel Matrix Algorithms and Applications.

\bibitem[Grigori and Tissot(2019)]{grigori2019scalable}
L.~Grigori and O.~Tissot.
\newblock Scalable linear solvers based on enlarged krylov subspaces with
  dynamic reduction of search directions.
\newblock \emph{SIAM Journal on Scientific Computing}, 41\penalty0
  (5):\penalty0 C522--C547, 2019.

\bibitem[Guermond et~al.(2021)Guermond, Maier, Popov, and Tomas]{Guermond21}
J.-L. Guermond, M.~Maier, B.~Popov, and I.~Tomas.
\newblock Second-order invariant domain preserving approximation of the
  compressible {N}avier--{S}tokes equations.
\newblock \emph{Computer Methods in Applied Mechanics and Engineering},
  375:\penalty0 113608, 2021.
\newblock \doi{10.1016/j.cma.2020.113608}.

\bibitem[Hager and Wellein(2011)]{Hager11}
G.~Hager and G.~Wellein.
\newblock \emph{Introduction to High Performance Computing for Scientists and
  Engineers}.
\newblock CRC Press, Boca Raton, 2011.

\bibitem[Kempf et~al.(2021)Kempf, He{\ss}, M\"{u}thing, and Bastian]{Kempf20}
D.~Kempf, R.~He{\ss}, S.~M\"{u}thing, and P.~Bastian.
\newblock Automatic code generation for high-performance discontinuous
  {G}alerkin methods on modern architectures.
\newblock \emph{{ACM} Transactions on Mathematical Software}, 47\penalty0
  (1):\penalty0 6:1--31, 2021.
\newblock \doi{10.1145/3424144}.

\bibitem[Kolev et~al.(2021)Kolev, Fischer, Min, Dongarra, Brown, Dobrev,
  Warburton, Tomov, Shephard, Abdelfattah, Barra, Beams, Camier, Chalmers,
  Dudouit, Karakus, Karlin, Kerkemeier, Lan, Medina, Merzari, Obabko, Pazner,
  Rathnayake, Smith, Spies, Swirydowicz, Thompson, Tomboulides, and
  Tomov]{Kolev2021}
T.~Kolev, P.~Fischer, M.~Min, J.~Dongarra, J.~Brown, V.~Dobrev, T.~Warburton,
  S.~Tomov, M.~S. Shephard, A.~Abdelfattah, V.~Barra, N.~Beams, J.-S. Camier,
  N.~Chalmers, Y.~Dudouit, A.~Karakus, I.~Karlin, S.~Kerkemeier, Y.-H. Lan,
  D.~Medina, E.~Merzari, A.~Obabko, W.~Pazner, T.~Rathnayake, C.~W. Smith,
  L.~Spies, K.~Swirydowicz, J.~Thompson, A.~Tomboulides, and V.~Tomov.
\newblock Efficient exascale discretizations: High-order finite element
  methods.
\newblock \emph{The International Journal of High Performance Computing
  Applications}, 35\penalty0 (6):\penalty0 527--552, 2021.
\newblock \doi{10.1177/10943420211020803}.

\bibitem[Krank et~al.(2017)Krank, Fehn, Wall, and Kronbichler]{Krank17}
B.~Krank, N.~Fehn, W.~A. Wall, and M.~Kronbichler.
\newblock A high-order semi-explicit discontinuous {G}alerkin solver for 3{D}
  incompressible flow with application to {DNS} and {LES} of turbulent channel
  flow.
\newblock \emph{Journal of Computational Physics}, 348:\penalty0 634--659,
  2017.
\newblock \doi{10.1016/j.jcp.2017.07.039}.

\bibitem[Kronbichler and Allalen(2018)]{Kronbichler18enviroinfo}
M.~Kronbichler and M.~Allalen.
\newblock Efficient high-order discontinuous {G}alerkin finite elements with
  matrix-free implementations.
\newblock In H.-J. Bungartz, D.~Kranzlm\"uller, V.~Weinberg, J.~Weism\"uller,
  and V.~Wohlgemuth, editors, \emph{Advances and New Trends in Environmental
  Informatics}, pages 89--110. Springer, 2018.
\newblock \doi{10.1007/978-3-319-99654-7_7}.

\bibitem[Kronbichler and Kormann(2012)]{Kronbichler11}
M.~Kronbichler and K.~Kormann.
\newblock A generic interface for parallel cell-based finite element operator
  application.
\newblock \emph{Computers and Fluids}, 63:\penalty0 135--147, 2012.
\newblock ISSN 0045-7930.
\newblock \doi{10.1016/j.compfluid.2012.04.012}.

\bibitem[Kronbichler and Kormann(2019)]{Kronbichler2019}
M.~Kronbichler and K.~Kormann.
\newblock Fast matrix-free evaluation of discontinuous {G}alerkin finite
  element operators.
\newblock \emph{ACM Transactions on Mathematical Software}, 45\penalty0
  (3):\penalty0 29:1--40, 2019.
\newblock \doi{10.1145/3325864}.

\bibitem[Kronbichler and Ljungkvist(2019)]{KronbichlerLjungkvist19}
M.~Kronbichler and K.~Ljungkvist.
\newblock Multigrid for matrix-free high-order finite element computations on
  graphics processors.
\newblock \emph{ACM Transactions on Parallel Computing}, 6\penalty0
  (1):\penalty0 2:1--32, 2019.
\newblock \doi{10.1145/3322813}.

\bibitem[Kronbichler and Wall(2018)]{Kronbichler18}
M.~Kronbichler and W.~A. Wall.
\newblock A performance comparison of continuous and discontinuous {G}alerkin
  methods with fast multigrid solvers.
\newblock \emph{SIAM Journal on Scientific Computing}, 40\penalty0
  (5):\penalty0 A3423--A3448, 2018.
\newblock \doi{10.1137/16M110455X}.

\bibitem[Kronbichler et~al.(2018)Kronbichler, Diagne, and
  Holmgren]{Kronbichler18multiphase}
M.~Kronbichler, A.~Diagne, and H.~Holmgren.
\newblock A fast massively parallel two-phase flow solver for microfluidic chip
  simulation.
\newblock \emph{The International Journal of High Performance Computing
  Applications}, 32\penalty0 (2):\penalty0 266--287, 2018.
\newblock \doi{10.1177/1094342016671790}.

\bibitem[Ljungkvist(2017)]{Ljungkvist17}
K.~Ljungkvist.
\newblock Matrix-free finite-element computations on graphics processors with
  adaptively refined unstructured meshes.
\newblock In \emph{HPC '17: Proceedings of the 25th High Performance Computing
  Symposium}, pages 1--12, San Diego, CA, USA, 2017. Society for Computer
  Simulation International.

\bibitem[Lockhart et~al.(2022)Lockhart, Bienz, Gropp, and
  Olson]{lockhart2022performance}
S.~Lockhart, A.~Bienz, W.~Gropp, and L.~Olson.
\newblock Performance analysis and optimal node-aware communication for
  enlarged conjugate gradient methods.
\newblock \emph{arXiv preprint arXiv:2203.06144}, 2022.

\bibitem[Malas et~al.(2017)Malas, Hager, Ltaief, and
  Keyes]{malas2017multidimensional}
T.~M. Malas, G.~Hager, H.~Ltaief, and D.~E. Keyes.
\newblock Multidimensional intratile parallelization for memory-starved stencil
  computations.
\newblock \emph{ACM Transactions on Parallel Computing}, 4\penalty0
  (3):\penalty0 12:1--32, 2017.
\newblock \doi{10.1145/3155290}.

\bibitem[MehriDehnavi et~al.(2013)MehriDehnavi, El-Kurdi, Demmel, and
  Giannacopoulos]{mehridehnavi2013communication}
M.~MehriDehnavi, Y.~El-Kurdi, J.~Demmel, and D.~Giannacopoulos.
\newblock Communication-avoiding {K}rylov techniques on graphic processing
  units.
\newblock \emph{IEEE transactions on magnetics}, 49\penalty0 (5):\penalty0
  1749--1752, 2013.
\newblock \doi{10.1109/TMAG.2013.2244861}.

\bibitem[Moxey et~al.(2020)Moxey, Amici, and Kirby]{Moxey20}
D.~Moxey, R.~Amici, and M.~Kirby.
\newblock Efficient matrix-free high-order finite element evaluation for
  simplicial elements.
\newblock \emph{{SIAM} Journal on Scientific Computing}, 42\penalty0
  (3):\penalty0 C97--C123, 2020.
\newblock \doi{10.1137/19m1246523}.

\bibitem[Munch et~al.(2021)Munch, Kormann, and Kronbichler]{munch2020hyper}
P.~Munch, K.~Kormann, and M.~Kronbichler.
\newblock hyper.deal: An efficient, matrix-free finite-element library for
  high-dimensional partial differential equations.
\newblock \emph{ACM Transactions on Mathematical Software}, 47\penalty0
  (4):\penalty0 33:1--34, 2021.
\newblock \doi{10.1145/3469720}.

\bibitem[Naumov(2016)]{naumov2016s}
M.~Naumov.
\newblock S-step and communication-avoiding iterative methods.
\newblock Technical Report NVR-2016-003, NVIDIA, 2016.

\bibitem[Orszag(1980)]{Orszag80}
S.~A. Orszag.
\newblock Spectral methods for problems in complex geometries.
\newblock \emph{Journal of Computational Physics}, 37\penalty0 (1):\penalty0
  70--92, 1980.
\newblock \doi{10.1016/0021-9991(80)90005-4}.

\bibitem[Patera(1984)]{Patera84}
A.~T. Patera.
\newblock A spectral element method for fluid dynamics: Laminar flow in a
  channel expansion.
\newblock \emph{Journal of Computational Physics}, 54\penalty0 (3):\penalty0
  468--488, 1984.
\newblock \doi{10.1016/0021-9991(84)90128-1}.

\bibitem[Rupp et~al.(2016)Rupp, Weinbub, J{\"u}ngel, and
  Grasser]{rupp2016pipelined}
K.~Rupp, J.~Weinbub, A.~J{\"u}ngel, and T.~Grasser.
\newblock Pipelined iterative solvers with kernel fusion for graphics
  processing units.
\newblock \emph{ACM Transactions on Mathematical Software}, 43\penalty0
  (2):\penalty0 11:1--27, 2016.
\newblock \doi{10.1145/2907944}.

\bibitem[Saad(1985)]{saad1985practical}
Y.~Saad.
\newblock Practical use of polynomial preconditionings for the conjugate
  gradient method.
\newblock \emph{SIAM Journal on Scientific and Statistical Computing},
  6\penalty0 (4):\penalty0 865--881, 1985.
\newblock \doi{10.1137/0906059}.

\bibitem[Solomonoff(1992)]{Solomonoff92}
A.~Solomonoff.
\newblock A fast algorithm for spectral differentiation.
\newblock \emph{J. Comput. Phys.}, 98\penalty0 (1):\penalty0 174--177, 1992.
\newblock \doi{10.1016/0021-9991(92)90182-X}.

\bibitem[Sun et~al.(2020)Sun, Mitchell, Kulkarni, Kl\"{o}ckner, Ham, and
  Kelly]{Sun20}
T.~Sun, L.~Mitchell, K.~Kulkarni, A.~Kl\"{o}ckner, D.~A. Ham, and P.~H. Kelly.
\newblock A study of vectorization for matrix-free finite element methods.
\newblock \emph{The International Journal of High Performance Computing
  Applications}, 34\penalty0 (6):\penalty0 629--644, 2020.
\newblock \doi{10.1177/1094342020945005}.

\bibitem[\'Swirydowicz et~al.(2019)\'Swirydowicz, Chalmers, Karakus, and
  Warburton]{Swirydowicz19}
K.~\'Swirydowicz, N.~Chalmers, A.~Karakus, and T.~Warburton.
\newblock Acceleration of tensor-product operations for high-order finite
  element methods.
\newblock \emph{The International Journal of High Performance Computing
  Applications}, 33\penalty0 (4):\penalty0 735--757, 2019.
\newblock \doi{10.1177/1094342018816368}.

\bibitem[{Treibig} et~al.(2010){Treibig}, {Hager}, and {Wellein}]{likwid10}
J.~{Treibig}, G.~{Hager}, and G.~{Wellein}.
\newblock {LIKWID}: A lightweight performance-oriented tool suite for x86
  multicore environments.
\newblock In \emph{2010 39th International Conference on Parallel Processing
  Workshops}, pages 207--216, 2010.
\newblock \doi{10.1109/ICPPW.2010.38}.

\bibitem[Tufo and Fischer(1999)]{Tufo99}
H.~M. Tufo and P.~F. Fischer.
\newblock Terascale spectral element algorithms and implementations.
\newblock In \emph{Proceedings of the 1999 ACM/IEEE conference on
  Supercomputing}, page~68. ACM, 1999.
\newblock \doi{10.1109/SC.1999.10035}.

\end{thebibliography}
\end{document}